\documentclass[prd,nofootinbib,superscriptaddress,preprint]{revtex4}
\usepackage[T1]{fontenc}
\usepackage{amsmath,amssymb}
\usepackage{epsfig}
\usepackage{graphicx,array}
\usepackage[usenames,dvipsnames]{color}
\usepackage{slashed}
\usepackage{comment}
\usepackage[colorlinks,citecolor=blue]{hyperref}
\usepackage{pdfpages}
\usepackage{color}
\usepackage{subfigure}

\usepackage{multirow}

\begin{document}

\title{\Large Discrete dark matter with light Dirac neutrinos} 
	\author{Debasish Borah}
	\email{dborah@iitg.ac.in}
	\affiliation{Department of Physics, Indian Institute of Technology, Guwahati, Assam 781039, India}
 \author{Pritam Das}
	\email{prtmdas9@gmail.com}
 \affiliation{Department of Physics, Salbari College, Baksa, Assam 781318, India}

\author{Biswajit Karmakar}
\email{biswajit.karmakar@us.edu.pl}
\affiliation{Institute of Physics, University of Silesia,  Katowice, Poland}
 	\author{Satyabrata Mahapatra}
	\email{satyabrata@g.skku.edu}
	\affiliation{Department of Physics and Institute of Basic Science, Sungkyunkwan University, Suwon 16419, Korea}

\begin{abstract}
We propose a new realisation of light Dirac neutrino mass and dark matter (DM) within the framework of a non-Abelian discrete flavour symmetry based on $A_4$ group. In addition to $A_4$, we also consider a $Z_2$ and an unbroken global lepton number symmetry $U(1)_L$ to keep unwanted terms away while guaranteeing the Dirac nature of light neutrinos. The field content, their transformations and flavon vacuum alignments are chosen in such a way that the type-I Dirac seesaw generates only one light Dirac neutrino mass while the other two masses arise from scotogenic contributions at one-loop. This leads to the Dirac scoto-seesaw framework, a generalisation of the widely studied scoto-seesaw model to Dirac neutrinos. The symmetry breaking of $A_4$ leaves a remnant $\mathcal{Z}_2$ symmetry responsible for stabilising DM. Dirac nature of light neutrinos introduces additional relativistic degrees of freedom $\Delta N_{\rm eff}$ within reach of cosmic microwave background experiments. 
\end{abstract}

\maketitle 

\section{Introduction}\label{sec:intro}
Origin of light neutrino mass and mixing \cite{ParticleDataGroup:2020ssz} has been one of the longstanding puzzles in particle physics. While neutrinos remain massless within the framework of the standard model (SM) of particle physics, several beyond the standard model (BSM) frameworks have been proposed to explain non-zero neutrino mass and mixing. Canonical seesaw frameworks \cite{Minkowski:1977sc, GellMann:1980vs, Mohapatra:1979ia,Sawada:1979dis,Yanagida:1980xy, Schechter:1980gr, Mohapatra:1980yp, Lazarides:1980nt, Wetterich:1981bx, Schechter:1981cv, Foot:1988aq} involve inclusion of heavy fields which couple to the SM leptons. These seesaw mechanisms and several of their descendants predict Majorana light neutrino masses, which violate the lepton number by two units. However, there has been no confirmation about the Majorana nature of light neutrinos with experiments searching for neutrinoless double beta decay (NDBD) continuing to report null results. While null results at NDBD experiments do not necessarily imply Dirac nature of light neutrinos, there has been growing interest in Dirac neutrino models of late. After a few initial attempts in this direction \cite{Roncadelli:1983ty, Roy:1983be, Babu:1988yq,
Peltoniemi:1992ss}, several new proposals to realise sub-eV scale Dirac neutrino 
masses~\cite{Chulia:2016ngi, Aranda:2013gga, 
Chen:2015jta, Ma:2015mjd, Reig:2016ewy, Wang:2016lve, Wang:2017mcy, Wang:2006jy, 
Gabriel:2006ns, Davidson:2009ha, Davidson:2010sf, Bonilla:2016zef, 
Farzan:2012sa, Bonilla:2016diq, Ma:2016mwh, Ma:2017kgb, Borah:2016lrl, 
Borah:2016zbd, Borah:2016hqn, Borah:2017leo, CentellesChulia:2017koy, 
Bonilla:2017ekt, Memenga:2013vc, Borah:2017dmk, CentellesChulia:2018gwr, 
CentellesChulia:2018bkz, Han:2018zcn, Borah:2018gjk, Borah:2018nvu, Borah:2019bdi, CentellesChulia:2019xky, Jana:2019mgj, Nanda:2019nqy, Guo:2020qin, Bernal:2021ezl, Borah:2022obi, Li:2022chc, Dey:2024ctx, Singh:2024imk} have appeared recently. Most of these works, in addition to the inclusion of new fields, also consider new discrete or continuous symmetries in order to forbid unwanted terms in the Lagrangian. In the spirit of the Dirac seesaw for light neutrinos, such unwanted terms include the bare mass term of Majorana type for singlet right handed neutrinos $(\nu_R)$ and direct coupling of $\nu_R$ with the SM lepton and Higgs doublets.

Another observed phenomenon which can not be explained in the SM is the presence of dark matter (DM) in the Universe. Its presence is supported by astrophysical observations at different scales \cite{Zwicky:1933gu, Rubin:1970zza, Clowe:2006eq} together with cosmological 
experiments like PLANCK, WMAP predicting around 26.8\% of the present Universe to be made up of DM \cite{ParticleDataGroup:2020ssz, Planck:2018vyg}. While weakly interacting massive particle (WIMP) has been the most widely studied particle DM framework, null results at direct detection experiments \cite{LUX-ZEPLIN:2022qhg} have also led to growing interest in alternative scenarios where DM is more feebly coupled to the SM, like the feebly interacting massive particle (FIMP). Recent reviews of WIMP and FIMP can be found in \cite{Arcadi:2017kky} and \cite{Bernal:2017kxu} respectively.

Motivated by this, we propose a new flavour symmetric origin of light Dirac neutrino mass and dark matter in this work. We consider the popular non-Abelian discrete flavour symmetry group of $A_4$ augmented with an additional discrete $Z_2$ and global lepton number $U(1)_L$ symmetries. The $A_4$ flavour symmetry gets spontaneously broken by flavon fields while generating light Dirac neutrino mass and mixing. For a review of non-Abelian discrete flavour symmetries and their consequences, see Ref.~\cite{Altarelli:2010gt, King:2013eh, Petcov:2017ggy,  Chauhan:2023faf, Ding:2024ozt} and references therein. Unlike popular DM scenarios with additional stabilising symmetries, here we have stable DM by virtue of a remnant $\mathcal{Z}_2$ symmetry naturally emerging from $A_4$ breaking. Similar scenario for Majorana light neutrino, coined as {\it discrete dark matter} was proposed in \cite{Hirsch:2010ru} and subsequently explored within different contexts \cite{Meloni:2010sk, Boucenna:2011tj, Adulpravitchai:2011ei, Eby:2011qa, Boucenna:2012qb, Hamada:2014xha, Lamprea:2016egz, DeLaVega:2018bkp, Bonilla:2023pna, Kumar:2024zfb}. {We propose a first realisation of this discrete dark matter paradigm with light Dirac neutrinos in a minimal $A_4$ flavour symmetric setup. We show that neutrino oscillation data can be satisfied after taking tree-level and one-loop contributions to light Dirac neutrino masses. Interestingly, this also leads to the first generalisation of the popular {\it Scoto-Seesaw} scenario \cite{Rojas:2018wym} to Dirac neutrinos, and can be termed as {\it Dirac Scoto-Seesaw}.} While the model predicts either scalar or fermion DM with the desired relic abundance, light Dirac neutrinos lead to observable $\Delta N_{\rm eff}$ at future cosmic microwave background (CMB) experiments like CMB Stage IV (CMB-S4) \cite{Abazajian:2019eic}, SPT-3G \cite{SPT-3G:2014dbx}, Simons Observatory \cite{SimonsObservatory:2018koc} and CMB-HD \cite{CMB-HD:2022bsz}.

This paper is organised as follows. In section \ref{sec:model}, we discuss our model and the origin of light Dirac neutrino mass with stable dark matter. In section \ref{sec:neutrino}, we discuss the numerical analysis related to neutrino mass and mixing followed by discussion of dark matter, $\Delta N_{\rm eff}$ and other detection prospects in section \ref{sec:dm}, \ref{sec:neff}, \ref{sec:lfv} respectively. Finally, we conclude in section \ref{sec:conclude}.

\section{The Model}\label{sec:model}
In this section, we discuss the relevant field content of the model, the field transformations under the chosen symmetries and the particle spectrum. Table \ref{table:A4 table} summarises the relevant field content and the corresponding quantum numbers. The SM lepton doublets and right chiral part of Dirac neutrino $(\nu_R)$ transform as $A_4$ singlets while heavy vector-like SM gauge singlet fermions $N_{L,R}$ transform as $A_4$ triplets. In addition to the SM Higgs $H$ transforming as $A_4$ singlet, additional scalar doublet $\eta$ and scalar singlet $\chi$ both transforming as $A_4$ triplets are included. The additional $Z_2$ symmetry prevents direct coupling of SM lepton doublet, SM Higgs doublet and $\nu_R$. On the other hand, a global unbroken lepton number symmetry $U(1)_L$ prevent Majorana mass terms from guaranteeing the Dirac nature of light neutrinos. {It should be noted that Dirac neutrino seesaw models typically require more symmetries and field content compared to the Majorana neutrino seesaw models due to the requirements of keeping lepton number conserved and the fact that Dirac neutrino has twice the degrees of freedom of its Majorana counterpart. Our choice of symmetry and field content is however not unique. One can replace global lepton number symmetry by  discrete symmetries like $Z_4$ \cite{CentellesChulia:2016rms} as well. On the other hand, $Z_2$ symmetry ensures seesaw origin of Dirac neutrino mass \cite{CentellesChulia:2016rms} in such a minimal setup. One can embed such additional discrete symmetries within gauge symmetries as well, but those scenarios will have more particle content. Here we stick to the most minimal extension of the discrete dark matter model \cite{Hirsch:2010ru} to realise light Dirac neutrinos.}

\begin{table}[h!]
		\begin{center}
			\begin{tabular}{c|ccccccccccc}
				\hline
				Fields & $\ell_{e}$ & $\ell_{\mu}$ & $\ell_{\tau}$ & $e_R$ & $ \mu_R$ & $\tau_R $ & $N_{L,R}$  & $\nu_{R_{e,\mu,\tau}}$ & $\eta$& $\chi$ & $H$\\
				\hline
                $SU(2)$ & 2 & 2 & 2 & 1 & 1 & 1 & 1 & 1&  2 &1 & 2 \\
				$A_4$ & $1$ & $1^{\prime\prime}$ & $1^{\prime}$ & $1$ & $1^{\prime \prime}$ & $1^{\prime}$ & $3$ & $1$ , $1'$ ,$1''$ & $3$ &$3$ & $1$\\
				$Z_2$ & $1$ & 1 & 1 & $1$ & $1$ & $1$ & $1$ & $-1$  & $1$& $-1$ & $1$\\
                $U(1)_L$ &  1 & 1 & 1 & 1 & 1 & 1 & 1 & 1 &  0 & 0 & 0 \\
			    \hline
			\end{tabular}
		\caption{Field contents and transformation under the symmetries of our model. }
		\label{table:A4 table}
		\end{center}
	\end{table}
 
The Yukawa Lagrangian for charged lepton can be written as
\begin{equation}
    -\mathcal{L}_{\rm CL} \supset Y_e \overline{\ell_e} H e_R + Y_\mu \overline{\ell_\mu} H \mu_R + Y_\tau \overline{\ell_\tau} H \tau_R + {\rm h.c.}
    \label{CLY}
\end{equation}
On the other hand, the neutral fermion Lagrangian is given by
\begin{eqnarray}\label{eq:Lag-nu2}
	-\mathcal{L}_{\nu} & \supset &{y_1}\bar{\ell_{e}}(N_R\Tilde{\eta})_1 +{y_2}\bar{\ell_{\mu}}(N_R\Tilde{\eta})_{1''}  + {y_3}\bar{\ell_{\tau}}(N_R\Tilde{\eta})_{1'}\nonumber  \\&~ & +{M_{N}(\Bar{N_L}N_R)_1}+y'_1(\bar{N}_L \chi)_1\nu_{Re}+y'_2(\bar{N}_L \chi)_{1''}\nu_{R\mu}+y'_3(\bar{N}_L \chi)_{1'}\nu_{R\tau} + {\rm h.c.}\\
 &=&{y_1}(\bar{\ell_{e}}N_{R_1}\Tilde{\eta_1}+\bar{\ell_{e}}N_{R_2}\Tilde{\eta_2}+\bar{\ell_{e}}N_{R_3}\Tilde{\eta_3}) +{y_2}(\bar{\ell_{\mu}}N_{R_1}\Tilde{\eta_1}+\omega\bar{\ell_{\mu}}N_{R_2}\Tilde{\eta_2}+\omega^2\bar{\ell_{\mu}}N_{R_3}\Tilde{\eta_3}) \nonumber  \\ &~ &+ {y_3}(\bar{\ell_{\tau}}N_{R_1}\Tilde{\eta_1}+\omega^2\bar{\ell_{\tau}}N_{R_2}\Tilde{\eta_2}+\omega\bar{\ell_{\tau}}N_{R_3}\Tilde{\eta_3})+M_N\Bar{N}_{L_1}N_{R_1}+M_N\Bar{N}_{L_2}N_{R_2}+M_{N}\Bar{N}_{L_3}N_{R_3}
  \nonumber \\ 
  &~ &+ y'_1 (\bar{N}_{L_1} \chi_1+\bar{N}_{L_2} \chi_2+\bar{N}_{L_3} \chi_3)\nu_{Re} + y'_2 (\bar{N}_{L_1} \chi_1+\omega\bar{N}_{L_2} \chi_2+\omega^2\bar{N}_{L_3} \chi_3)\nu_{R\mu} \nonumber \\ 
  &~ &+  y'_3 (\bar{N}_{L_1} \chi_1+\omega^2\bar{N}_{L_2} \chi_2+\omega\bar{N}_{L_3} \chi_3)\nu_{R\tau} + {\rm h.c.}
  \label{eq:lag2}
 \end{eqnarray}
where the $A_4$ product rules, given in Appendix \ref{apa}, have been used to write the interactions in component form. The scalar potential of the model is given in Appendix \ref{appen2}. The Yukawa Lagrangian for charged leptons given by Eq. \eqref{CLY} gives rise to a diagonal charged lepton mass matrix. 
\begin{figure}[h]
    \centering
    \includegraphics[scale=0.15]{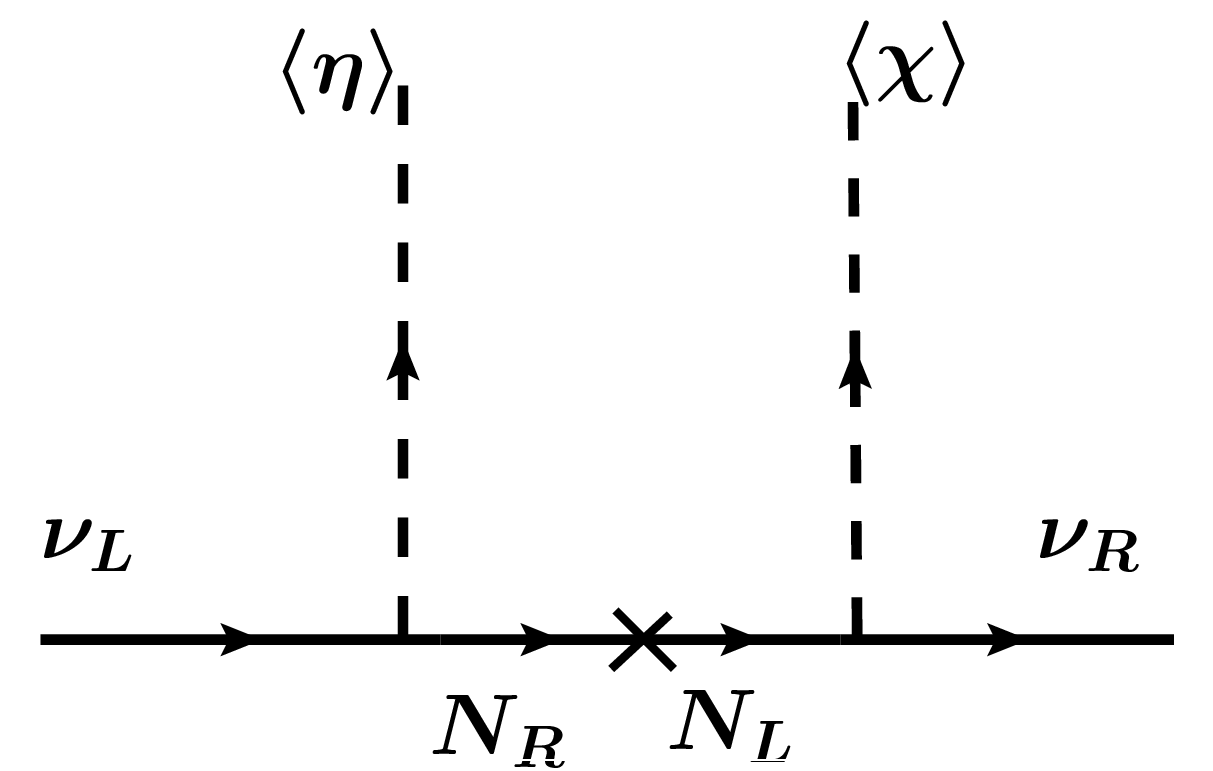}
\includegraphics[scale=0.17]{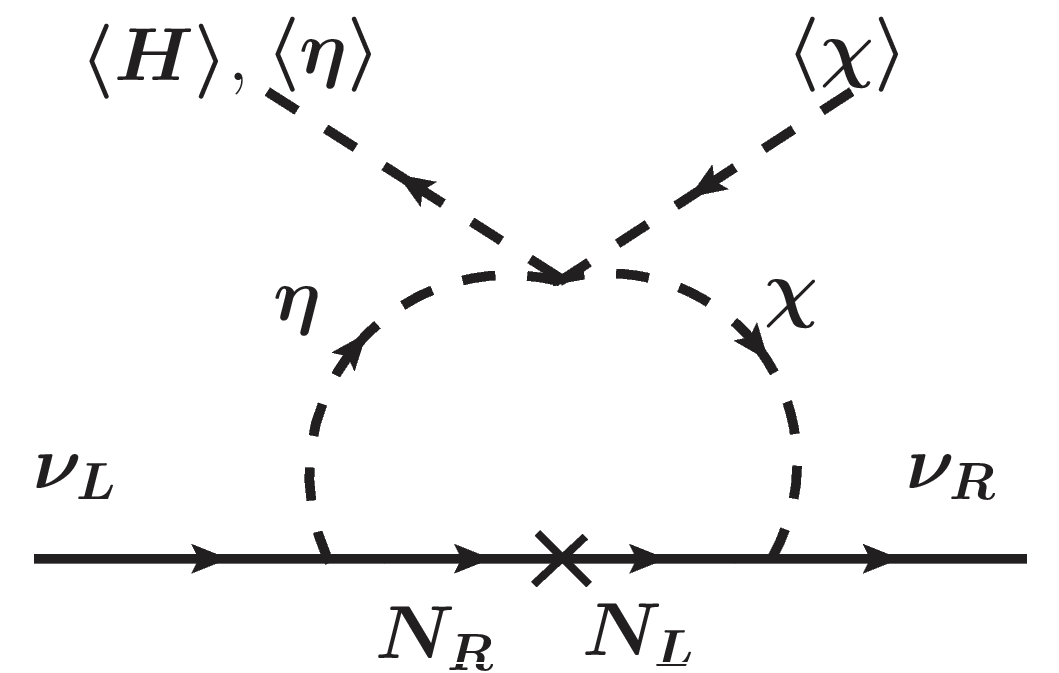}
    \caption{Neutrino mass generation at tree level and one-loop level.}
    \label{fig:1}
\end{figure}
At tree level the neutral fermion mass matrices in the ($\nu_L,N_R$), ($N_L, \nu_R$), $(N_L,N_R)$ basis can be written as 
\begin{eqnarray}
	m_D= \frac{v_{\eta}}{\sqrt{2}}\begin{pmatrix}
		y_1 & 0 & 0 \\
		y_2 & 0 & 0 \\
		y_3 & 0 & 0
	\end{pmatrix}, m'_D= \frac{v_{\chi}}{\sqrt{2}}\begin{pmatrix}
		y'_1 & y'_2 & y'_3 \\
		0 & 0 & 0 \\
		0 & 0 & 0
	\end{pmatrix}, 
 M_R= \begin{pmatrix}
		M_N & 0 & 0 \\
		0 & M_N  & 0 \\
		0 & 0 & M_N 
	\end{pmatrix}; 
\end{eqnarray} 
respectively, where the vacuum expectation values (VEVs) of the $A_4$ scalar triplets are considered to be $\langle \eta \rangle=(\langle \eta_1 \rangle, \langle \eta_2 \rangle, \langle \eta_3 \rangle)=(\frac{v_\eta}{\sqrt{2}},0,0)$ and $\langle \chi \rangle=(\langle \chi_1 \rangle, \langle \chi_2 \rangle, \langle \chi_3 \rangle)=(\frac{v_\chi}{\sqrt{2}},0,0)$. Denoting the VEV of the neutral component of the Higgs doublet $H$ as $v$, we have the constraint $\sqrt{v^2_\eta +v^2}=246$ GeV from the requirement of desired electroweak symmetry breaking (EWSB). Therefore, at three level, via type-I seesaw like scenario (shown in the left panel of Fig. \ref{fig:1}), the light Dirac neutrino mass matrix  can be written as
\begin{eqnarray} \label{eq:type-I}
m_{\nu 0}=-m_D M^{-1}_R m'_D=-\frac{v_{\eta}v_{\chi}}{2M_N}\begin{pmatrix}
		y_1y'_1 & y_1y'_2 & y_1y'_3 \\
		y_2y'_1 & y_2y'_2 & y_2y'_3 \\
		y_3y'_1 & y_3y'_2 & y_3y'_3
	\end{pmatrix}.
\end{eqnarray}
Clearly, the tree level neutrino mass matrix is a matrix of rank-1 predicting only one massive neutrino with mass $m_0=-\frac{v_{\eta}v_{\chi}}{2M_N}(y_1y'_1+y_2y'_2+ y_3y'_3)$. 

At one-loop, all the $A_4$ triplet scalars $\eta=(\eta_1,\eta_2,\eta_3)$ and $\chi=(\chi_1,\chi_2,\chi_3)$ irrespective of their VEVs can contribute to light neutrino mass, as shown in the right panel of Fig. \ref{fig:1}. This contribution is similar to the Dirac scotogenic\footnote{Scotogenic model was originally proposed in \cite{Tao:1996vb, Ma:2006fn} for Majorana neutrinos.} model discussed in several earlier works \cite{Gu:2007ug, Farzan:2012sa, Borah:2016zbd, Ma:2016mwh, Borah:2017leo, Wang:2017mcy, Ma:2019iwj, Ma:2019yfo, Ma:2019coj, Leite:2020wjl, Guo:2020qin, Bernal:2021ezl, Chowdhury:2022jde, Borah:2022phw, Borah:2022enh}. The one-loop contribution to light Dirac neutrino mass can be estimated as 
\begin{eqnarray}
m^{\rm loop}_{\nu } = \sum^3_{j=1} Y_{kj} Y'_{jl}  M_j \mathcal{F}_j ,
\end{eqnarray}
where  
\begin{eqnarray}\label{eq:loop-fn}
\mathcal{F}_j=\sum_{k,l} \frac{\sin\theta_{kl} \cos\theta_{kl}}{32 \pi^2}\left( \frac{m^2_{\xi_k}}{m^2_{\xi_k}-M^2_j}{\rm ln}\frac{m^2_{\xi_k}}{M^2_j}-\frac{m^2_{\xi_l}}{m^2_{\xi_l}-M^2_j}{\rm ln}\frac{m^2_{\xi_l}}{M^2_j} \right),
\end{eqnarray}
with $m_{\xi_k}, m_{\xi_l}$ being the mass eigenstates of the associated scalars, $\theta_{kl}$ being the mixing angle and $M_j=M_N$. Following Eq. \eqref{eq:lag2}, the Yukawa matrices $Y$, $Y'$ involved in the scotogenic contribution above can be written as 
\begin{eqnarray}
    Y=\begin{pmatrix}
		y_1   &  y_1  & y_1 \\
		y_2 &  \omega y_2 & \omega^2 y_2 \\
		y_3 &  \omega^2 y_3 & \omega y_3
	\end{pmatrix}, ~~ Y'=\begin{pmatrix}
		y'_1   &  y'_2 & y'_3 \\
		y'_1 &  \omega y'_2 & \omega^2 y'_3 \\
		y'_1 &  \omega^2 y'_2 & \omega y'_3
	\end{pmatrix}. 
\end{eqnarray}
Using these, the complete one-loop mass matrix for light Dirac neutrinos can be written as 
\begin{eqnarray}
m^{\rm loop}_{\nu }  & = & Y_{k1}Y'_{1l}M_N\mathcal{F}_1+Y_{k2}Y'_{2l}M_N\mathcal{F}_2+Y_{k3}Y'_{3l}M_N\mathcal{F}_3\\
&=& M_N\begin{pmatrix}
		y_1y'_1(\mathcal{F}_1+\mathcal{F}_2+\mathcal{F}_3)   &  y_1y'_2(\mathcal{F}_1+\omega\mathcal{F}_2+\omega^2\mathcal{F}_3) & y_1y'_3(\mathcal{F}_1+\omega^2\mathcal{F}_2+\omega\mathcal{F}_3) \\
		y_2y'_1(\mathcal{F}_1+\omega\mathcal{F}_2+\omega^2\mathcal{F}_3) & y_2y'_2(\mathcal{F}_1+\omega^2\mathcal{F}_2+\omega\mathcal{F}_3) & y_2y'_3(\mathcal{F}_1+\mathcal{F}_2+\mathcal{F}_3)  \\
		y_3y'_1(\mathcal{F}_1+\omega^2\mathcal{F}_2+\omega\mathcal{F}_3) &  y_3y'_2(\mathcal{F}_1+\mathcal{F}_2+\mathcal{F}_3) & y_3y'_3(\mathcal{F}_1+\omega\mathcal{F}_2+\omega^2\mathcal{F}_3)
	\end{pmatrix}\nonumber\\\label{eq:dirac-scoto}
\end{eqnarray}
Therefore, the light Dirac neutrino mass matrix, taking tree level and one-loop contributions, can be written as 
\begin{eqnarray}\label{eq:mtotal}
m_{\nu}&=&m_{\nu 0}+m^{\rm loop}_{\nu }= \begin{pmatrix}
           (m_{\nu})_{11}  &  (m_{\nu})_{12}  & (m_{\nu})_{13}\\
		(m_{\nu})_{21}&  (m_{\nu})_{22} & (m_{\nu})_{23}\\
		(m_{\nu})_{31} &  (m_{\nu})_{32} & (m_{\nu})_{33}
    \end{pmatrix} 
\end{eqnarray}
with the details of the components $(m_{\nu})_{ij}$ being given in Appendix \ref{appen3}. For simplicity, the Yukawa couplings appearing in the neutrino mass matrix are considered to be real. Thus, CP violation is introduced only through the complex parameter $\omega$,  appearing here only in the radiative contribution as a consequence of the considered $A_4$ discrete flavour symmetry.  This mass matrix has rank 3 and predicts three non-zero mass eigenvalues, in general. Also, it does not possess any specific symmetric structure and hence can be fitted to neutrino mixing data by appropriate choice of parameters, as we discuss below. Interestingly, in the limit $y_2=y_3$ and $y'_2=y'_3$, the effective mass matrix given in Eq. (\ref{eq:mtotal}) yields $\mu-\tau$ reflection symmetric mass matrix~\cite{Harrison:2002et} predicting fixed values for $\theta_{23} (=\pi/4)$ and $\delta_{\rm CP}(=\pm \pi/2$), see Appendix \ref{appen3} for details. Since type-I seesaw generates only one mass scale with the other being generated by the scotogenic contribution, our framework provides a generalisation of the scoto-seesaw mechanism \cite{Rojas:2018wym, Mandal:2021yph, Barreiros:2020gxu, Barreiros:2022aqu, Ganguly:2022qxj, Ganguly:2023jml, Kumar:2023moh} to light Dirac neutrinos. Similar to earlier works on Majorana neutrinos, the Dirac scoto-seesaw framework can explain the two mass scales in the neutrino sector namely, atmospheric and solar mass scales, from tree level (type-I seesaw) and radiative (scotogenic) contributions, respectively.

It should be noted that the above choice of VEV alignment leaves a remnant $\mathcal{Z}_2 \subset A_4$ unbroken after spontaneous symmetry breaking. The following fields transform non-trivially under this remnant symmetry as 
\begin{equation}
    \mathcal{Z}_2: \,\, \eta_{2,3} \rightarrow -\eta_{2,3}; \,\,\,\, \chi_{2,3} \rightarrow -\chi_{2,3}; \,\,\,\, N_{2,3} \rightarrow -N_{2,3}
\end{equation}
while all other fields are $\mathcal{Z}_2$-even. Therefore, the lightest $\mathcal{Z}_2$-odd particle is stable and can be a dark matter candidate, the details of which are discussed in section \ref{sec:dm}.

\section{Fit to Neutrino data}
\label{sec:neutrino}

The effective Dirac neutrino mass matrix obtained in our framework is constituted by both tree level type-I Dirac seesaw contribution as well as one-loop Dirac scotogenic contributions as shown in Fig.~\ref{fig:1}. The effective neutrino mass matrix, given in Eq.~(\ref{eq:mtotal}), is dependent on the associated Yukawa couplings, scalar and fermion mass parameters, and mixing of the scalar mass eigenstates.  Owing to the considered discrete flavour symmetry, here, both Dirac type-I seesaw and scotogenic contributions share the same Yukawa couplings.  Hence, as a consequence,  the Yukawa couplings $y_i, y'_i$ both appear in these two contributions,  given in Eq.~(\ref{eq:type-I}) and (\ref{eq:dirac-scoto}) respectively. Furthermore, same pair of these Yukawa couplings emerges in both Dirac type-I seesaw and scotogenic contributions in each element of the effective neutrino mass matrix, see Appendix \ref{appen3} for details. {Since heavy fermion and new scalar masses are kept in the TeV ballpark for desired dark matter phenomenology, the product of the Yukawa couplings are restricted to be small as we will see below. This is in sharp contrast with low scale seesaw models where only radiative contribution is present or with high scale type-I seesaw model where Yukawa couplings can be of $\mathcal{O}(1)$.} 
The procured light neutrino mass matrix $m_{\nu}$ takes the most general form, having the ability to reproduce correct neutrino mixing corresponding to a wide range of the associated parameters.  Therefore to explain the observed neutrino oscillation data~\cite{Esteban:2020cvm,deSalas:2020pgw}, for simplicity, we consider three benchmark cases for the Yukawa couplings $y_i, y'_i$ such as:  Case-A when   $y_i \gg y'_i$, Case-B when $y_i \ll y'_i$ and Case-C when $y_i \simeq y'_i$ respectively. Based on these classifications of the Yukawa couplings, to reproduce correct light neutrino mass, we restrict the flavon VEVs to be around or above the electroweak scale while the $M_N$ is found to lie in the multi-TeV range from the fitting. In each of these cases, we consider the 3$\sigma$ allowed range of neutrino oscillation 
data~\cite{Esteban:2020cvm} on the mixing angles $\theta_{12}$, $\theta_{23}$, $\theta_{13}$, mass-squared differences $\Delta m^2_{21}$, $\Delta m^2_{31}$ and their ratio $\Delta m^2_{21}/\Delta m^2_{31}$ to obtain the constraints on the parameters appearing in the light Dirac neutrino mass matrix. 
\begin{figure}[h!]
	\begin{center}
		\includegraphics[width=.32\textwidth]{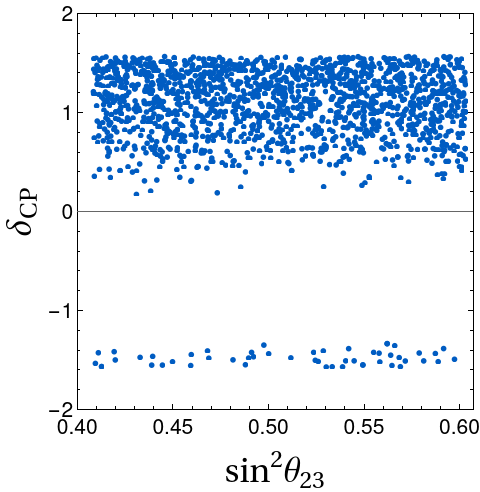}
        \includegraphics[width=.32\textwidth]{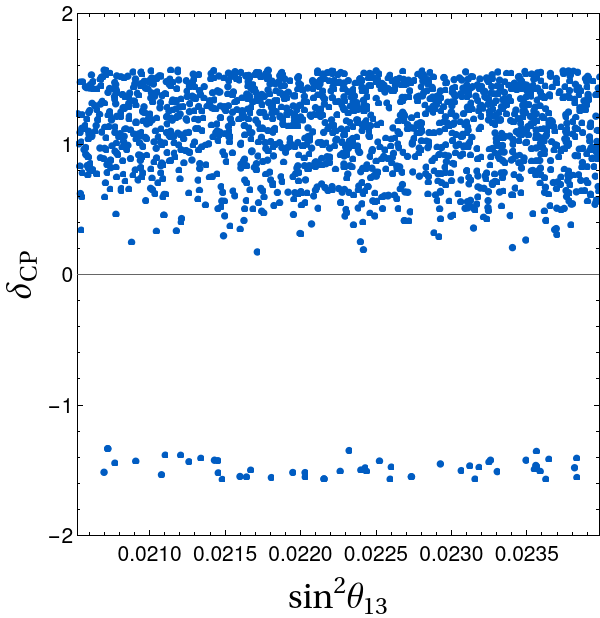}
        
		\includegraphics[width=.32\textwidth]{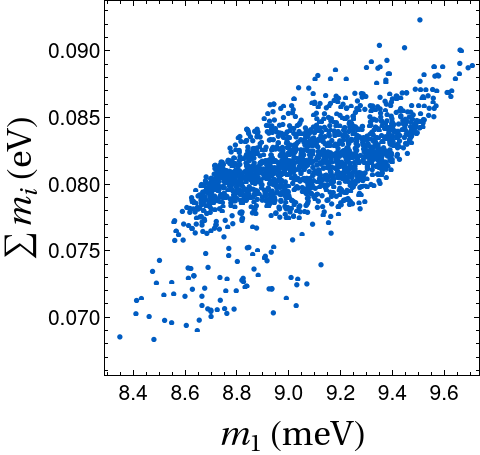}
		\includegraphics[width=.32\textwidth]{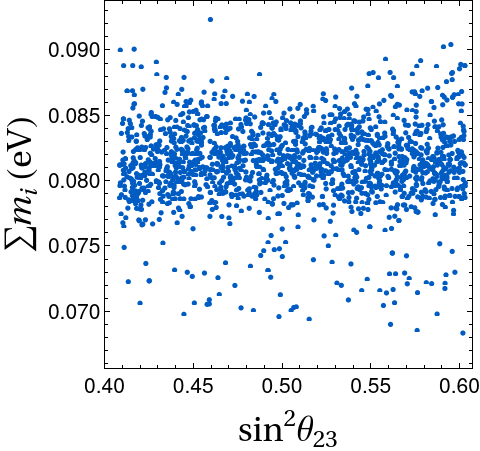}
\end{center}
\caption{Correlation  for $\delta_{\rm CP}-\sin^2\theta_{23}$, $\delta_{\rm CP}-\sin^2\theta_{13}$, $\sum m_i-m_1$, $\sum m_i-\sin^2\theta_{23}$  for $y_i \gg y'_i$ (Case-A).}
\label{fig:caseA-para-cor}
\end{figure}
In the subsequent sections, we
will explore the viability of these three benchmark scenarios to obtain dark matter relic density, direct search constraints and $\Delta N_{\rm eff}$ elucidating a flavour symmetric origin for these observables.   For brevity here we explicitly show the analysis for the normal hierarchy, however such an analysis can also be obtained 
for inverted hierarchy of the neutrino masses.

{\bf Case-A: }
We first consider the limiting case $y_i \gg y'_i$ for the associated Yukawa couplings. Using the the 3$\sigma$ range of neutrino oscillation 
data~\cite{Esteban:2020cvm}, 
$y'_i$ are found to be in the range $6 \times 10^{-11}\leq y'_1 \leq 10^{-10}$, $3 \times 10^{-11}\leq y'_{2,3} \leq 8.5 \times 10^{-11}$
and corresponding ranges for $y_i$ are found to be $0.03\leq y_1 \leq 0.05 $, $0.052\leq y_2 \leq 0.1 $ and $0.05\leq y_3 \leq 0.1 $ respectively. To be in agreement with neutrino oscillation data, other parameters 
are restricted within the range $2.8 \, \leq (M_N/{\rm TeV}) \leq 5$, $20\leq (v_{\eta}/{\rm GeV}) \leq 240$, $0.22 \leq (v_{\chi}/{\rm TeV}) \leq 1.8$ with the loop functions defined in Eq.~(\ref{eq:loop-fn}) having the magnitude $\mathcal{F}_i \sim \mathcal{O}(10^{-4})$.  With the estimation of the parameters appearing in the neutrino mass matrix, we are now in a position to show the correlation among neutrino oscillation parameters as shown in Fig.~\ref{fig:caseA-para-cor}. In this figure,  we have plotted the correlations for $\delta_{\rm CP}-\sin^2\theta_{23}$, $\delta_{\rm CP}-\sin^2\theta_{13}$, $\sum m_i-m_1$, $\sum m_i-\sin^2\theta_{23}$ planes for Case-A and  predicted ranges for $\delta_{\rm CP}$, $\sum m_i$ and $m_1$ are found to be in the range $(0.15-1.6), -(1.3-1.6)$ radian, $(0.06-0.09)$ eV and  $(8.4-9.7)$ meV respectively.

{\bf Case-B: } We now consider the benchmark scenario, $y_i \ll y'_i$. Using a similar methodology,  we can again constrain the Yukawa couplings $y_i, y'_i$, imposing the constraints on neutrino masses and mixing angles~\cite{Esteban:2020cvm}. 
The Yukawa couplings $y_i, y'_i$ are found to be restricted within the range $1.4\times 10^{-11}\leq y_1 \leq 2.5 \times 10^{-11}$, $2\times 10^{-11}\leq y_2 \leq  7 \times 10^{-11}$, $3.4\times 10^{-11} \leq  y_3\leq 5.5 \times 10^{-11}$
\begin{figure}[h]
	\begin{center}
		\includegraphics[width=.325\textwidth]{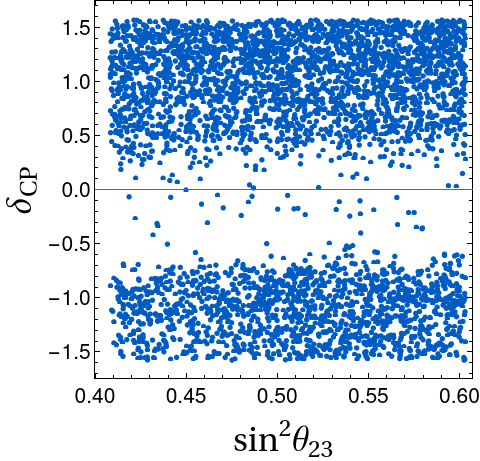}
        \includegraphics[width=.310\textwidth]{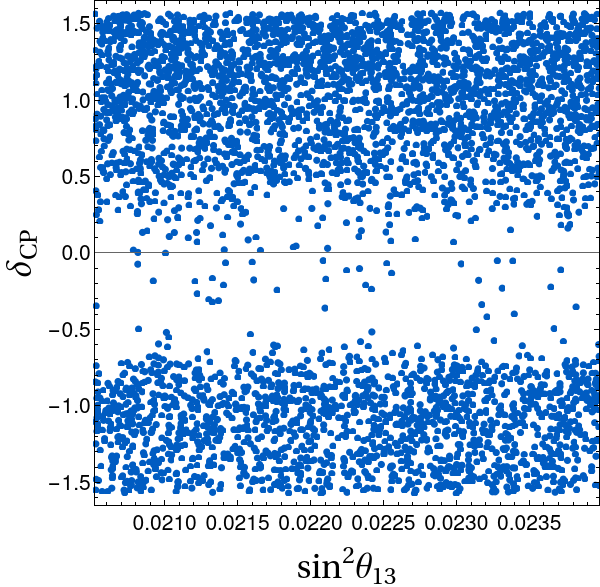}

		\includegraphics[width=.32\textwidth]{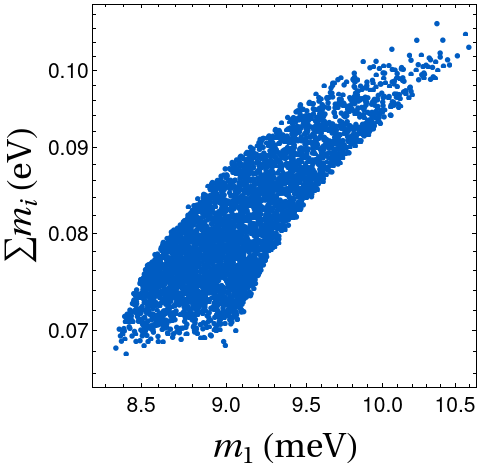}
		\includegraphics[width=.32\textwidth]{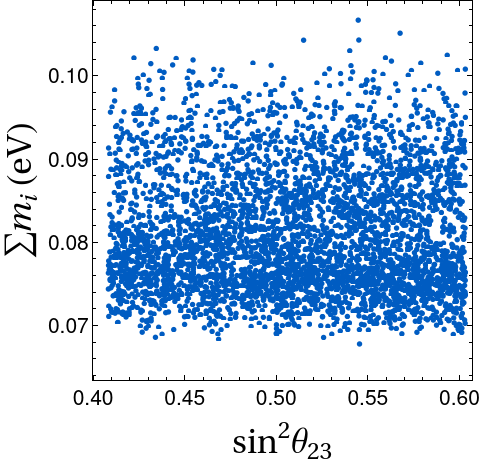}
\end{center}
\caption{Correlation  for $\delta_{\rm CP}-\sin^2\theta_{23}$, $\delta_{\rm CP}-\sin^2\theta_{13}$, $\sum m_i-m_1$ and $\sum m_i-\sin^2\theta_{23}$  for  $y_i \ll y'_i$ (Case-B).}
\label{fig:caseB-para-cor}
\end{figure}
and $0.01\leq y'_1\leq 0.1$, $0.03\leq y'_2\leq 0.1$, $0.04\leq y'_3\leq 0.1$ respectively. 
In order to satisfy the neutrino oscillation data, other parameters are restricted within the range $6 \,  \leq (M_N/{\rm TeV}) \leq 10$, $20\leq (v_{\eta}/{\rm GeV}) \leq 240$, $0.2 \leq (v_{\chi}/{\rm TeV}) \leq 1.8$. Corresponding loop functions $\mathcal{F}_i$,  defined in Eq.~(\ref{eq:loop-fn}), appearing in the Dirac scotogenic contribution are found to be in the range $5\times 10^{-4} \leq \mathcal{F}_1 \leq 7\times 10^{-4}$, $3\times 10^{-4} \leq \mathcal{F}_2 \leq  10^{-3}$,  $8\times 10^{-4} \leq \mathcal{F}_3 \leq 1.3\times 10^{-3}$.  After estimating the parameters relevant to the light neutrino mass matrix, we proceed to find some correlations among neutrino oscillation parameters as shown in Fig.~\ref{fig:caseB-para-cor}. In this figure,  we have plotted the correlations for $\delta_{\rm CP}-\sin^2\theta_{23}$, $\delta_{\rm CP}-\sin^2\theta_{13}$, $\sum m_i-m_1$, $\sum m_i-\sin^2\theta_{23}$ planes for the scenario $y_i \ll y'_i$ and  predicted ranges for $\delta_{\rm CP}$, $\sum m_i$ and $m_1$ are found to be in the range $-1.6 \leq (\delta_{\rm CP}/{\rm radian}) \leq  1.6$, $(0.068-0.11)$ eV and  $(8.4-10.5)$ meV respectively.

{\bf Case-C:}
Finally, we now consider the third benchmark scenario where $y_i \simeq y'_i$. In this scenario, one can also satisfy the observed constraints on neutrino masses and mixing  
where the associated Yukawa couplings are found to be in the range $3 \times 10^{-7}\leq y_1 \leq 4.6 \times 10^{-7}$, $6.9 \times 10^{-7}\leq y_2 \leq 7.9\times 10^{-7}$, $7.1 \times 10^{-7}\leq y_3 \leq 8.1 \times 10^{-7}$
and $5.6\times 10^{-6}\leq y'_1 \leq 6.8\times 10^{-6} $, $2.9\times 10^{-6}\leq y'_2 \leq 5.5\times 10^{-6} $ and $3.6\times 10^{-6}\leq y'_3 \leq 4.75\times 10^{-6} $ respectively.
\begin{figure}[h]
	\begin{center}
		\includegraphics[width=.32\textwidth]{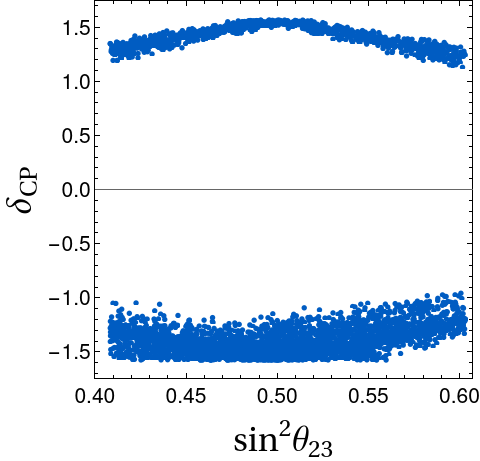}
        \includegraphics[width=.31\textwidth]{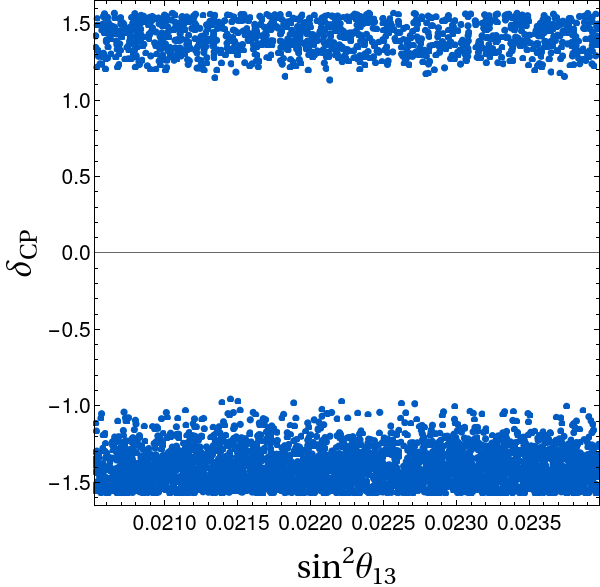}

		\includegraphics[width=.32\textwidth]{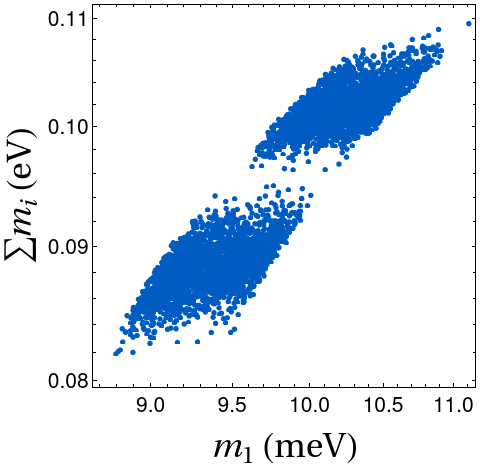}
		\includegraphics[width=.325\textwidth]{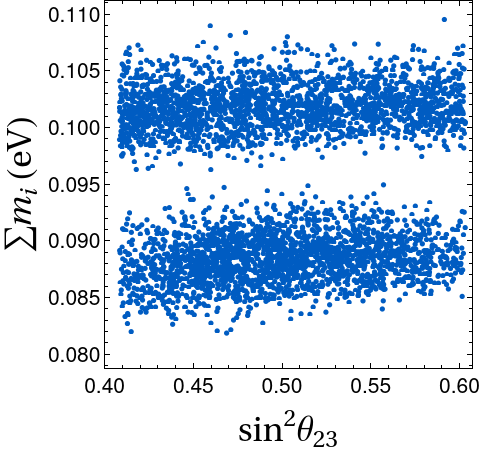}
\end{center}
\caption{Correlation  for $\delta_{\rm CP}-\sin^2\theta_{23}$, $\delta_{\rm CP}-\sin^2\theta_{13}$, $\sum m_i-m_1$ and $\sum m_i-\sin^2\theta_{23}$  for $y_i \simeq y'_i$ (Case-C).}
\label{fig:caseC-para-cor}
\end{figure}
To fit the model with the neutrino oscillation data in this scenario, other parameters are restricted within the range $8 \, \leq (M_N/{\rm TeV} ) \leq 16$, $35\leq (v_{\eta}/{\rm GeV}) \leq 240$, $0.23 \leq (v_{\chi}/{\rm TeV}) \leq 1.8$. Similarly,  loop functions $\mathcal{F}_i$  defined in Eq.~(\ref{eq:loop-fn}), appearing in the Dirac scotogenic contribution are found to be in the range $10^{-3} \leq \mathcal{F}_1 \leq 1.2\times 10^{-3}$, $4\times 10^{-4} \leq \mathcal{F}_2 \leq 7\times 10^{-4}$,  $10^{-4} \leq \mathcal{F}_3 \leq 1.7\times 10^{-4}$.  Similar to the other two cases, here also we show the correlation among neutrino oscillation parameters in Fig.~\ref{fig:caseC-para-cor}. In this figure,  we have plotted the correlations in $\delta_{\rm CP}-\sin^2\theta_{23}$ and $\delta_{\rm CP}-\sin^2\theta_{13}$ in the upper plane, while $\sum m_i-m_1$, $\sum m_i-\sin^2\theta_{23}$ in the lower planes and the predicted ranges for $\delta_{\rm CP}$, $\sum m_i$ and $m_1$ are found to be in the range $1.2 \leq (\delta_{\rm CP}/{\rm radian}) \leq  1.58, -1.58 \leq (\delta_{\rm CP}/{\rm radian}) \leq  -1.0$, $\sum m_i \in (0.082-0.11)$ eV and  $m_1 \in (8.8-11)$ meV respectively.

It is worth mentioning that the product of the two Yukawa couplings $y_i y'_j$ remain around $\lesssim \mathcal{O}(10^{-12})$ in all the cases mentioned above. This is particularly due to the chosen scale of seesaw $\sim \mathcal{O}(10)$ TeV, singlet flavon VEV around or above the electroweak scale and fine-tuning in the loop function at the level of $\mathcal{O}(10^{-4})$. Unlike in canonical seesaw models where Dirac Yukawa couplings can be large for a very high seesaw scale, scoto-seesaw models offer somewhat less freedom to do so as the same heavy particle masses and Dirac Yukawa couplings take part in tree-level and one-loop contribution to seesaw masses. In the spirit of the scoto-seesaw mechanism both of these contributions are needed to satisfy neutrino data as tree-level seesaw alone gives a neutrino mass matrix of rank-1 only. We check that any fine-tuning of Yukawa couplings can be totally avoided at the expense of fine-tuning the loop function to $\lesssim \mathcal{O}(10^{-14})$. For a given seesaw scale, such fine-tuned loop function can be achieved by choosing tiny scalar quartic couplings involved in $\eta-\chi$ interactions. We, however, do not explore such possibilities. Additionally, the scale of seesaw around a few tens of TeV is chosen keeping in mind the upper limit on thermal dark matter mass which we will discuss in the next section.

In the following sections, for compactness,  we will consider these three benchmark scenarios for studies on scalar/fermionic  WIMP dark matter, $\Delta N_{\rm eff}$ and LFV decays associated with this discrete dark matter model for Dirac neutrinos. 

\section{Dark matter}
\label{sec:dm}

In this model, gauge singlet Dirac fermions $N_{2,3}$, gauge singlet scalars $\chi_{2,3}$ and $SU(2)$ doublet scalars $\eta_{2,3}$, remain odd under the residual $\mathcal{Z}_2$ symmetry such that the lightest among them becomes a viable dark matter candidate. When considering the fermionic dark matter scenario, $i.e.$, $M_N < M_\eta < M_\chi$, achieving the correct relic density becomes challenging because the mass $M_N$ is constrained to the multi-TeV range by the neutrino mass criteria. Fermion DM in the scotogenic model was studied in several earlier works; see \cite{Ahriche:2017iar, Borah:2018smz, Mahanta:2019gfe} for example. As can be seen from these earlier works, the fermion singlet DM relic relies primarily on Yukawa portal annihilations and coannihilations. Due to the involvement of the same Yukawa couplings in both type-I and scotogenic contributions to neutrino mass, we can not have Yukawa coupling in $ \geq \mathcal{O}(1) $ regime as discussed in the previous section. After including coannihilation among fermion DM $N$ and $\mathcal{Z}_2$-odd scalars, correct relic abundance can be found only for Case-A with $3000\le (M_N/\text{GeV}) \le5000$, considering the parameter space in agreement with neutrino data. However, for Case-B with $ 6000\le (M_N/\text{GeV})\le10000$ and Case-C with $11400\le (M_N/\text{GeV})\le 14000$, where $M_N$ is pushed to even larger values, it is not possible to achieve the correct relic density for the fermionic dark matter via thermal freeze-out. This is because the large $M_N$ value and insufficient (co)annihilation rates always lead to an overabundant relic density. The thermal relic of $N$ in Case-A can be obtained due to smaller $M_N$ and large co-annihilation rate with $SU(2)$ doublet scalars via large Yukawa portal ($y_i \gg y'_i$) interactions. Since doublet scalars have more efficient annihilation rates, it is easier to obtain the correct relic of $N$ in Case-A compared to Case-B, where $N$ has sizeable Yukawa with scalars which are predominantly singlets. While Case-C offers the possibility of non-thermal or freeze-in fermion singlet DM, we find it to be overproduced as well leaving only Case-A as viable for fermion DM scenario.

\begin{figure}[h]
   \centering
   \includegraphics[scale=0.5]{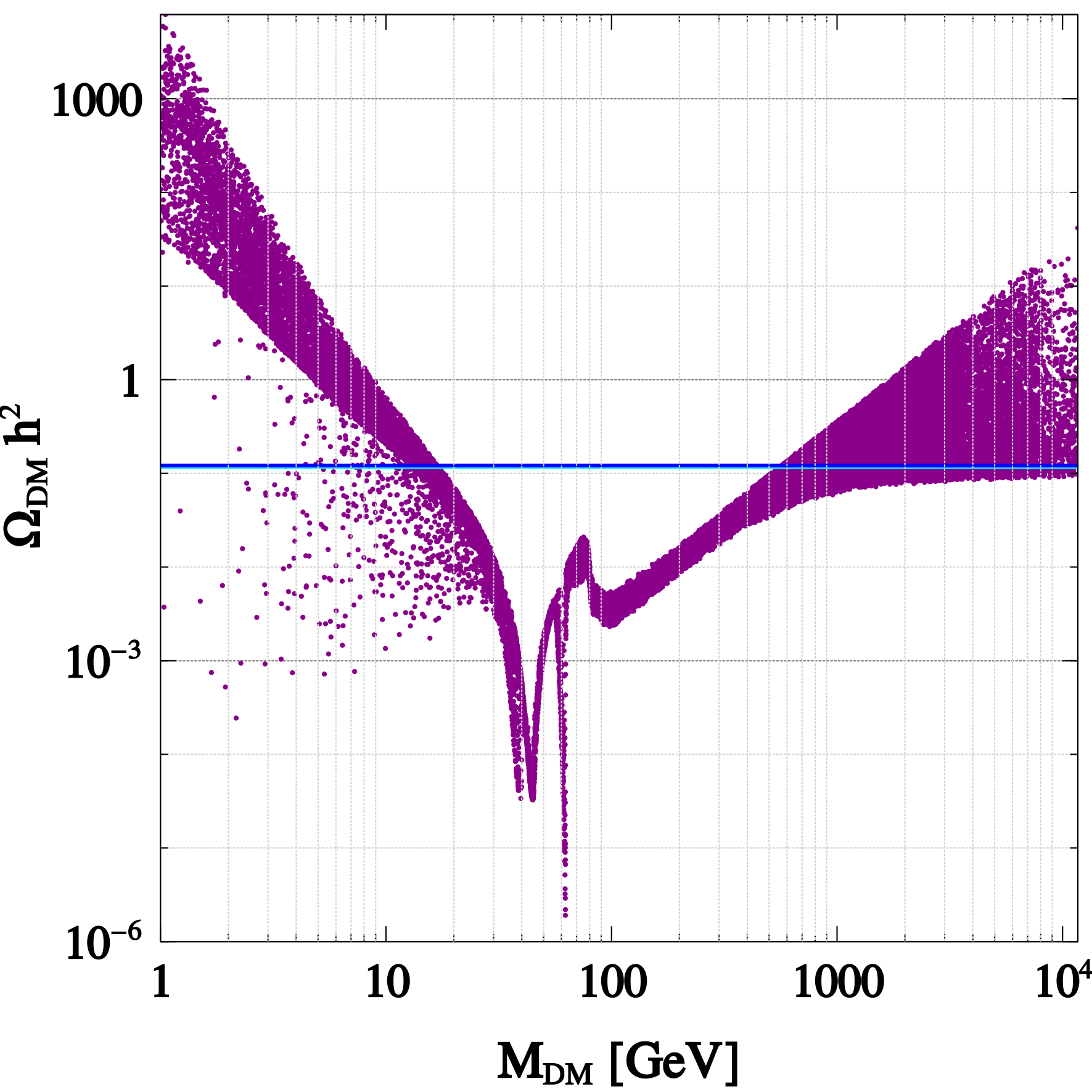}
    \includegraphics[scale=0.5]{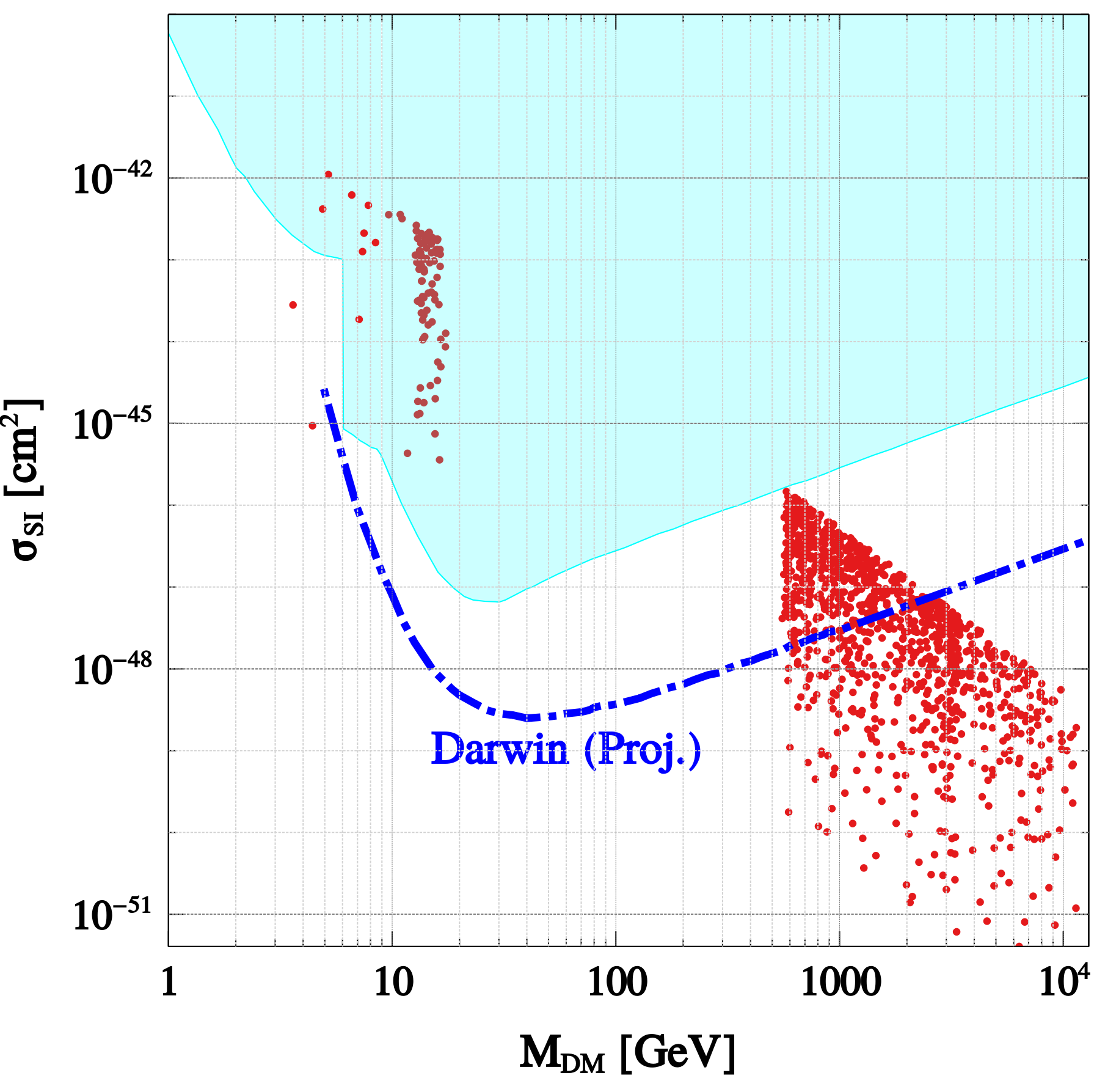}
    \caption{{\it Left panel:} variation of scalar DM relic abundance with its mass. The blue solid line indicates the observed DM relic. {\it Right panel:} spin-independent DM-nucleon scattering as a function of the DM mass. The red coloured points satisfy DM relic. The cyan-shaded region is ruled out by direct detection constraints.  }
\label{fig:dmrelicdd}
\end{figure}

In the case of scalar DM, the correct relic density can be obtained in all three cases discussed in the previous section due to the gauge and scalar portal annihilation and co-annihilation of the scalar DM while successfully explaining the neutrino data. Due to the possibility of additional portals namely gauge and scalar portals for annihilation, scalar DM relic can be obtained without relying on the Yukawa portal, unlike fermion singlet DM. With $M_N$ varied in the required range so as to satisfy the neutrino oscillation data, we vary the masses of $\mathcal{Z}_2$-odd scalars $\zeta_i$ (see Appendix \ref{appen2} for the details of physical scalar masses and eigenstates) assuming two hierarchies $M_{\zeta_i}<M_{N}$ or $M_{\zeta_1}<M_{N}<M_{\zeta_j}, (j \neq 1)$. We consider $\zeta_1$, the lightest $\mathcal{Z}_2$-odd scalar to be the DM. As the mass difference between $\zeta_i$'s are constrained in a specific range from the neutrino mass criteria when the mass hierarchy is $M_{\zeta_i}<M_{N}$, it is possible to get DM masses in the sub-TeV range. However when we assume the hierarchy $M_{\zeta_1}<M_{N}<M_{\zeta_j} (j \neq 1)$, the DM mass always lies in the multi-TeV scale. 

In the left panel of Fig.~\ref{fig:dmrelicdd}, we show the variation of scalar DM relic with its mass. As shown in several earlier works on scalar DM, the (co)annihilations mediated by electroweak gauge bosons and the SM Higgs are responsible for the dip for intermediate mass, leaving two distinct mass ranges satisfying the correct relic. The right panel of the same figure shows the spin-independent DM-nucleon scattering cross-section (mediated via the Higgs portal) as a function of scalar DM mass for the points which satisfy relic abundance criteria. We used the package \texttt{MicrOMEGAs} \cite{Alguero:2023zol} to calculate the relic density and DM-nucleon scattering cross-section after generating the model files using \texttt{LanHEP} \cite{Semenov:2014rea}. The cyan-shaded region is ruled out by the most stringent direct search constraints from LZ-2022~\cite{LUX-ZEPLIN:2022qhg}  for heavier DM and CRESST-III~\cite{CRESST:2019jnq} and DarkSide-50~\cite{DarkSide:2018bpj} for lighter DM. Clearly, we can see that the viable DM mass range that satisfies the correct relic density and direct detection rate ranges from around $500$ GeV to several TeV scales. Interestingly, future direct detection experiment Darwin~\cite{DARWIN:2016hyl} can probe a part of this high mass region up to $M_{\rm DM} \sim 3$ TeV.

In the discussions above, we have kept DM mass upto 10 TeV firstly due to the availability of direct detection prospects and secondly due to the fact that thermal DM gets overproduced if we go up by one order of magnitude $M_{\rm DM} \gtrsim \mathcal{O}(100)$ TeV, referred to as the unitarity bound \cite{Griest:1989wd, Bhatia:2020itt}. While such strict model-independent upper bound does not apply to non-thermal DM, we find such DM (fermion singlet) to be overproduced even with masses within tens of TeV, as mentioned earlier. Increasing fermion singlet DM masses beyond tens of TeV will further increase relic due to proportionate increase in Yukawa couplings from seesaw criteria and the fact that non-thermal DM yield increases with increasing interaction rate \cite{Hall:2009bx}.



 \section{Observable $\Delta N_{\rm eff}$}
 \label{sec:neff}
 The effective number of relativistic degrees of freedom ${\rm N}_{\rm eff}$ defined as 
\begin{equation}
    {\rm N}_{\rm eff} = \frac{8}{7} \left ( \frac{11}{4} \right)^{4/3} \left ( \frac{\rho_{R}-\rho_{\gamma}}{\rho_{\gamma}} \right )\,,
\end{equation}
where $\rho_{R}, \rho_{\gamma}$ denote total radiation and photon densities respectively. Existing data from CMB experiments like PLANCK constrains such additional light species by putting limits on the effective
degrees of freedom for neutrinos during the
era of recombination ($z\sim 1100$) as  \cite{Planck:2018vyg} 
\begin{eqnarray}
{\rm
N_{eff}= 2.99^{+0.34}_{-0.33}
}
\label{Neff}
\end{eqnarray}
at $2\sigma$ or $95\%$ C.L. including baryon acoustic oscillation (BAO) data. At $1\sigma$ C.L. it becomes more stringent to ${\rm N}_{\rm eff} = 2.99 \pm 0.17$. A similar bound also exists from big bang nucleosynthesis (BBN), $2.3 < {\rm N}_{\rm eff} <3.4$ at $95\%$ C.L. \cite{Cyburt:2015mya}. Recent observations by the DESI collaboration report $\Delta {\rm N}_{\rm eff} < 0.4$ at $2\sigma$ \cite{DESI:2024mwx}, which is weaker than the PLANCK bound given in Eq. \eqref{Neff}. All these bounds are consistent with SM predictions ${\rm N^{SM}_{eff}}=3.045$ \cite{Mangano:2005cc, Grohs:2015tfy,deSalas:2016ztq}. Future CMB experiment CMB-S4 is expected reach a much better sensitivity of $\Delta {\rm N}_{\rm eff}={\rm N}_{\rm eff}-{\rm N}^{\rm SM}_{\rm eff}
= 0.06$ \cite{Abazajian:2019eic}, taking it closer to the SM prediction. Another future CMB experiment, specially, CMB-HD \cite{CMB-HD:2022bsz} can probe $\Delta {\rm N}_{\rm eff}$ up to $0.014$ at $1\sigma$. Dirac nature of light neutrinos introduces additional relativistic degrees of freedom in the form of right chiral neutrinos, which can contribute to $\Delta {\rm N}_{\rm eff}$ either thermally or non-thermally. Light Dirac neutrino models often lead to enhanced $\Delta {\rm N}_{\rm eff}$, some recent works on which can be found in \cite{Abazajian:2019oqj, FileviezPerez:2019cyn, Nanda:2019nqy, Han:2020oet, Luo:2020sho, Borah:2020boy, Adshead:2020ekg, Luo:2020fdt, Mahanta:2021plx, Du:2021idh, Biswas:2021kio, Borah:2022obi, Li:2022yna, Das:2023yhv}. Depending upon the strength of Yukawa couplings associated with $\nu_R$, we can have either thermal or non-thermal contribution to $\Delta {\rm N}_{\rm eff} $ as we discuss in the following subsections.

\subsection{Thermal $\Delta {\rm N}_{\rm eff}$}
 \begin{figure}[h]
    \centering
    \includegraphics[scale=0.5]{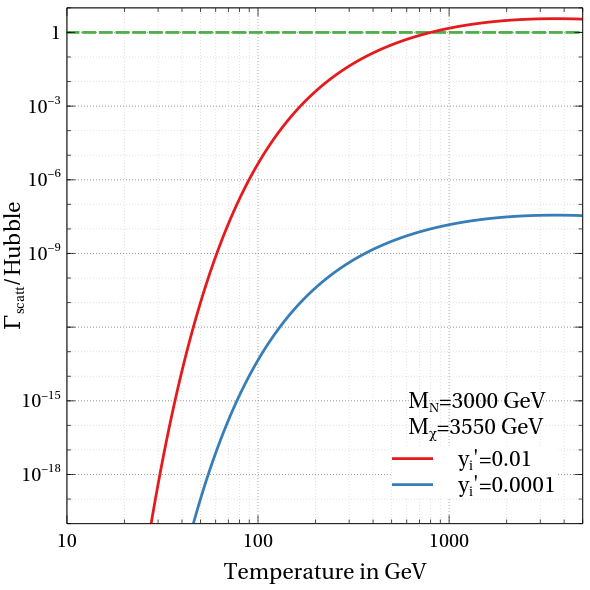}
    \caption{Interaction rate of $\nu_R$ compared to the Hubble expansion rate, indicating its decoupling profile for two sets of benchmark values.}
    \label{decP}
\end{figure}
We have three generations of right handed neutrinos ($\nu_{R_i}$) in our model, and their abundance in the Universe will be decided by the interaction strength and the masses of the particle with which they interact. The last three interaction terms of the Lagrangian (Eq. \eqref{eq:Lag-nu2}) are the only three interactions that $\nu_R$ have with $N$ and $\chi$. The relative strength between the two types of Yukawa couplings $y_i$ and $y_i'$ leads to three distinctive classes (Case-A, Case-B, Case-C) as we discussed before, in the context of neutrino mass and DM. With sufficiently large Yukawa coupling ($y_i'$) as in Case-B, $\nu_R$ may enter the thermal bath with the SM degrees of freedom. As the Universe expands and the temperature falls, the scattering processes that kept $\nu_R$ in the thermal bath eventually go out of equilibrium, leading to the decoupling of $\nu_R$. For the combined scattering processes of $N_L\bar{\nu_R}\leftrightarrow N_L\bar{\nu_R}$ and $\chi\bar{\nu_R}\leftrightarrow \chi\bar{\nu_R}$, we check the decoupling or thermalization profile with two sets of benchmark parameters (masses and couplings) as shown in Fig. \ref{decP}. As $N$ and $\chi$ masses lie at $\mathcal{O}$(TeV) range, it leads to the decoupling of $\nu_R$ from the thermal bath very early. Such early decoupling, above the electroweak scale, leads to a fixed value $\Delta {\rm N}_{\rm eff}=0.14$ for three generations of $\nu_R$. Clearly, this remains within reach of several future CMB experiments mentioned earlier.

 \subsection{Non-thermal $\Delta {\rm N}_{\rm eff}$}
It is also possible to have observable $\Delta {\rm N}_{\rm eff}$ when $\nu_R$ does not enter the thermal bath. In such a case, $\nu_R$ can be produced non-thermally from the decay of the scalar $\chi$ as $\chi\rightarrow N \nu_{R_i}$ or from the decay of $N$ as $N \rightarrow \chi \nu_{R_i}$ depending upon our choice of fermion or scalar DM. Similar contributions can also arise from scatterings. With small Yukawa couplings for the $\bar{N_L}\chi\nu_{R}$ interaction (as in Case-A and Case-C), $\nu_R$ can not enter the thermal bath via the $N_L\bar{\nu_R}\leftrightarrow N_L\bar{\nu_R}$ (or $\chi\bar{\nu_R}\leftrightarrow \chi\bar{\nu_R}$) scattering process. In such a situation, where the scalar $\chi$ is in the thermal bath, $\nu_R$ are produced via the decay $\chi\rightarrow N\nu_R$ (assuming fermion DM scenario). 

 With $\chi$ in equilibrium and subsequently producing $N$ and $\nu_R$, the relevant Boltzmann equation for the comoving density of $\nu_R$ of flavour $\alpha$ is given by \cite{Biswas:2022vkq},
 \begin{eqnarray}
     \frac{d\tilde{Y}_{\nu^\alpha_R}}{dx}=\frac{\beta}{x \mathcal{H} s^{1/3}}\langle E\Gamma\rangle Y^{\rm eq}_{\chi}.
     \label{neffcasec}
 \end{eqnarray}
Here, $x=\frac{M_\chi}{T}$, $\mathcal{H}$ is the Hubble parameter in the radiation dominated era and $\beta$ is defined as $\beta=\big[1+T\frac{dg_s/dT}{3g_s}\big]$ with $g_s$ being the effective relativistic entropy degrees of freedom. For the sub-eV scale $\nu_R$, which remains relativistic till recombination, the comoving energy density is defined as $\tilde{Y}_{\nu^\alpha_R}=\rho_{\nu^\alpha_R}/s^{4/3}$ with $\rho_{\nu^\alpha_R}, s$ denoting energy density of $\nu^\alpha_R$ and entropy density of the Universe respectively. On the right hand side of Eq. \eqref{neffcasec}, $Y^{\rm eq}_{\chi} \equiv n^{\rm eq}_\chi/s$ denotes the equilibrium comoving density of $\chi$. The quantity $\langle E \Gamma \rangle$ is defined as
\begin{equation}
    \langle E \Gamma \rangle =\frac{M_{\chi}}{2}
    \left(1-\frac{M^2_{N}}{M^2_{\chi}}\right)\Gamma_{\chi}
\end{equation}
where $\Gamma_{\chi}$ denotes the partial decay width for the process $\chi \rightarrow N \nu_R$. The $\nu_R$ produced from $\chi$ decay will contribute to this additional radiation density. Therefore, with $\rho_{R}=\rho_{\rm SM}+\rho_{\nu_R}$, $\Delta {\rm N}_{\rm eff}$ for one species of $\nu_R$ (indicated by flavour index $\alpha$) calculated at the CMB epoch $(T_{\rm CMB})$ is
\begin{eqnarray}\label{neff1}
\nonumber    \Delta {\rm N}^\alpha_{\rm eff}&=&{\rm N}^\alpha_{\rm eff}-{\rm N}_{\rm eff}^{\rm SM}\\
    \nonumber&=&2 \times\Big(\frac{\rho_{\nu^\alpha_R}}{\rho_{\nu_L}}\Big)_{T_{\rm CMB}}\\
    \nonumber &=&2\times \Big(\frac{\rho_{\nu^\alpha_R}}{\rho_{\nu_L}}\Big)_{T_{\rm BBN}}\\
     &=&2\times \Big(\frac{s^{4/3}\tilde{Y}_{\nu^\alpha_R}}{\rho_{\nu_L}}\Big)_{T_{\rm BBN}}
\end{eqnarray}
where, the extrapolation from the BBN epoch $(T_{\rm BBN})$ to CMB epoch $(T_{\rm CMB})$ is justified by the fact that both $\rho_{\nu_L}, \rho_{\nu_R}$ redshift similarly for $T < T_{\rm BBN}$. One can simply calculate the energy density ratio at BBN epoch $\sim \mathcal{O}({10})$ MeV as $\nu_R$ production stops much earlier. The numerical factor 2 in Eq. \eqref{neff1} indicates the degrees of freedom of each $\nu_R$. The net contribution to $\Delta {\rm N}_{\rm eff}$ can be obtained by summing over flavour index $\alpha$ as $\Delta {\rm N}_{\rm eff} = \sum_\alpha \Delta {\rm N}^\alpha_{\rm eff}$. Ignoring the temperature dependence of the relativistic degrees of freedom, one can solve Eq. \eqref{neffcasec} analytically to find $\Delta {\rm N}_{\rm eff}$ \cite{Biswas:2022vkq}.
\begin{figure}
    \centering
    \includegraphics[scale=0.6]{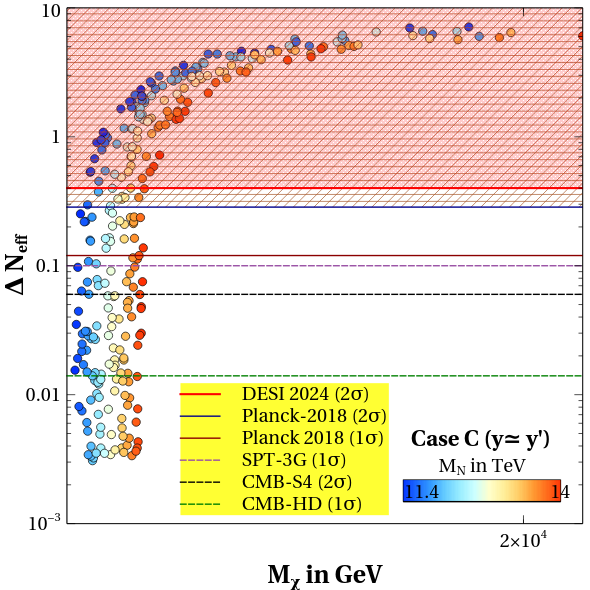}
    \caption{Contribution to non-thermal $\Delta N_{\rm eff}$ for the Case-C where $y_i'\sim \mathcal{O}(10^{-6})$,  considering the decay $\chi\rightarrow\nu_R N$.}
    \label{neffc}
\end{figure}

The decoupling profile for $\nu_R$ for two different benchmark choices of Yukawa coupling $y_i'$ is shown in Fig. \ref{decP}. For scalar ($\chi$) and fermion ($N$) masses in the TeV ballpark, we find that for $y_i'\le \mathcal{O}(10^{-3})$, the interaction $N_L\bar{\nu_R}\leftrightarrow N_L\bar{\nu_R}$ can not attain thermal equilibrium in the early Universe. Therefore, Case-A with $y_i'\sim\mathcal{O}(10^{-11})$, $\nu_R$ will never enter the thermal bath, hence, thermal $\Delta {\rm N}_{\rm eff}$ is not possible. On the other hand, $\nu_R$ produced non-thermally from the decay of either $\chi$ or $N$ gives rise to negligible contribution to $\Delta {\rm N}_{\rm eff}$. As mentioned before, the Case-B, with $0.01\le y_1'\le 0.1$,  $0.03\le y_2'\le 0.1$, $0.03\le y_3'\le 0.1$, we have thermal contribution to $\Delta {\rm N}_{\rm eff}$. Due to the decoupling of $\nu_R$ much above the electroweak scale, we have a fixed contribution $\Delta {\rm N}_{\rm eff}=0.14$ for Case-B. Finally, in Case-C, Yukawa couplings are of the size $\mathcal{O}(10^{-6})$, and the corresponding interactions do not allow $\nu_R$ to enter the thermal bath. Considering the decay of $\chi\rightarrow\nu_R N$, we have shown the non-thermal contributions to $\Delta N_{\rm eff}$ in Fig. \ref{neffc}. Clearly, some part of the parameter space is already ruled out by PLANCK 2018 and DESI 2024 limits while most of the remaining parameter space remains within reach of future CMB experiments like SPT-3G, CMB-S4 and CMB-HD. As there are multiple $\mathcal{Z}_2$-odd scalars, it is possible to have $\chi_i \rightarrow\nu_R N$ type decays from one of the heavier scalars irrespective of fermion or scalar DM. While contribution to $\Delta {\rm N}_{\rm eff}$ from scalar decay can be calculated simply by considering the scalar $\chi_i$ to be in equilibrium, contributions from the decay $ N\rightarrow\nu_R \chi_j$ will be more involved as the number density of non-thermal $N$ needs to be found by solving the respective Boltzmann equation \cite{Biswas:2022vkq}.


To summarise, observable $\Delta {\rm N}_{\rm eff}$ is possible in Case-B and Case-C. However, fermion DM is not possible in these two cases irrespective of the freeze-out or freeze-in mechanism. On the other hand, fermion DM produced via freeze-out mechanism is viable in Case-A even though the $\Delta {\rm N}_{\rm eff}$ remains out of reach from future CMB experiments. This conflict between observable $\Delta {\rm N}_{\rm eff}$ and DM relic can be avoided if we consider scalar DM. Scalar DM allows light scalars in the sub-TeV ballpark with observable direct detection rates while keeping collider prospects more promising compared to the fermion DM scenario where all $\mathcal{Z}_2$-odd particles remain in the multi-TeV ballpark.

It should be noted that we have calculated DM relic and $\Delta N_{\rm eff}$ for the points shown in Fig. \ref{fig:caseA-para-cor}, \ref{fig:caseB-para-cor}, \ref{fig:caseC-para-cor}. Since some of these points are disallowed by relic and $\Delta N_{\rm eff}$ constraints, shown in Fig. \ref{fig:dmrelicdd} and Fig. \ref{neffc} respectively, it is possible, in principle, to rule out some of the points in Fig. \ref{fig:caseA-para-cor}, \ref{fig:caseB-para-cor}, \ref{fig:caseC-para-cor} as well. However, as mentioned while describing the fit to neutrino data, we considered the entire loop function as a parameter instead of the details of scalar masses and mixing. On the other hand DM and $N_{\rm eff}$ phenomenology consider specific scalar DM masses satisfying those loop functions. Therefore, implementing the constraints from DM and $\Delta N_{\rm eff}$ on neutrino data allowed points in Fig. \ref{fig:caseA-para-cor}, \ref{fig:caseB-para-cor}, \ref{fig:caseC-para-cor} will involve simultaneous generation of the complete scalar and fermion singlet mass spectra which we have not performed during our numerical scan. While a complete parameter scan using all phenomenological bounds is beyond the scope of the present work, we show the interplay of different bounds and observables highlighting the detection prospects of this Dirac scoto-seesaw model.
\section{Other detection prospects}
\label{sec:lfv}
\begin{figure}[h!]
    \centering
        \includegraphics[scale=0.6]{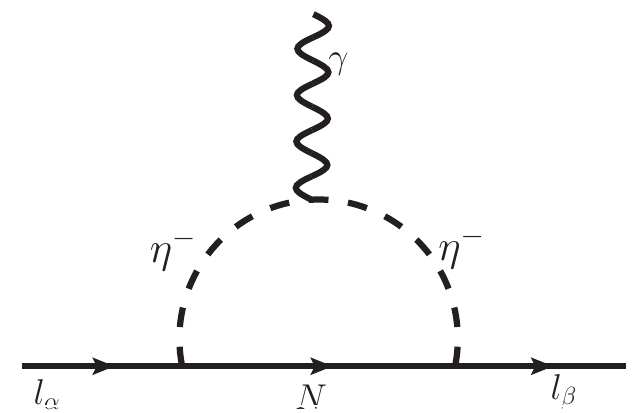}
    \caption{One-loop contribution to charged lepton flavour violating $\ell_\eta\rightarrow\ell_\alpha\gamma$ process }
    \label{lfv1}
\end{figure}
In addition to the detection possibilities of DM and $\Delta {\rm N}_{\rm eff}$, the model can have other detection aspects too. We briefly comment on such possibilities in this section. 

Firstly, we can have a lepton flavour violating (LFV) process at one-loop level, as shown in Fig. \ref{lfv1}. The corresponding decay width can be calculated using the prescriptions given in \cite{Kuno:1999jp, Guo:2020qin}. With relevant Yukawa couplings of $y_i \sim \mathcal{O}(10^{-1}-10^{-2})$, as in Case-A, it gives the branching ratio $\text{Br}(\mu\rightarrow e\gamma)\sim\mathcal{O}(10^{-15})$, which is close to the recent upper bound indicated by MEG-II experiment $\text{Br}(\mu\rightarrow e\gamma)<3.1\times 10^{-13}$ \cite{MEGII:2023ltw}. For Case-B and Case-C with smaller Yukawa couplings $y_i \sim \mathcal{O}(10^{-11})$ and $y_i \sim \mathcal{O}(10^{-7})$, we get very small branching ratio $\text{Br}(\mu\rightarrow e\gamma)$ as $\mathcal{O}(10^{-57})$ and $\mathcal{O}(10^{-41})$ respectively, beyond experimental reach in near future.

Due to the presence of several new scalars, admixtures of singlet and doublet, we can have different collider prospects as well. If scalar DM mass $(M_\chi)$ is lighter than the SM Higgs mass $(M_h)$ such that $h \rightarrow \chi \chi$ decay is allowed, it can contribute to invisible Higgs decay. Such invisible Higgs decay is currently constrained at Br$(h \rightarrow {\rm invisible}) < 0.14$ \cite{ATLAS:2022yvh}. Production of $\mathcal{Z}_2$-odd scalars in colliders like the large hadron collider (LHC) can give rise to pure leptonic final states plus missing transverse energy (MET) \cite{Gustafsson:2012aj, Datta:2016nfz}, hadronic final states plus MET \cite{Poulose:2016lvz} or a mixture of both. Such MET can arise from light Dirac neutrinos or scalar/fermion DM in our scenario. It is also possible to have more exotic signatures like tri-lepton plus MET \cite{Hashemi:2016wup} and mono-jet \cite{Belyaev:2016lok, Belyaev:2018ext}. It should be noted that we also have new scalars in the model which are $\mathcal{Z}_2$-even. They can lead to additional final states, similar to generic two-Higgs doublet models. The new charged scalars can also modify the diphoton decay width of the SM Higgs, currently constrained at $\frac{{\rm Br}(h\rightarrow \gamma \gamma)_{\rm expt}}{{\rm Br}(h\rightarrow \gamma \gamma)_{\rm SM}}=1.12\pm0.09$ \cite{CMS:2021kom}.

Another interesting detection prospect arises in the gravitational wave (GW) frontier. Spontaneous breaking of discrete symmetries like $A_4 \times Z_2 $, as we have in our model, can lead to the formation of topological defects like domain walls (DW). While DW, if stable or allowed to dominate the energy density of the Universe, will be in conflict with cosmological observations, they can be made to disappear by introducing small bias, in the form of explicit symmetry breaking terms. Such annihilating or collapsing DW then generates stochastic GW which can be probed at ongoing or future experiments. GW signatures of minimal Dirac seesaw model was studied in earlier works \cite{Barman:2022yos, Barman:2023fad} while similar studies in the context of flavour symmetric models like $A_4$ can be found in \cite{Gelmini:2020bqg}. The bias or explicit symmetry breaking terms introduced for DW disappearance may also lead to interesting indirect detection signatures of DM due to its decay into different standard model particles\footnote{See \cite{King:2023ayw, Borah:2024kfn} for related discussions.}. While the model has several detection prospects, it can also be falsified by future observation of neutrinoless double beta decay.

\begin{table}[h!]
\begin{centering}
    \begin{tabular}{|c||c|c|c|c|c|c|c|}
         \hline
         &Yukawa&Neurino& Observable $\Delta N_{\rm eff}$ &Observable $\Delta N_{\rm eff}$ & Fermion & Scalar &Observable\\
         Case&Coupling& Data & (Thermal)  & (Non-thermal) & DM& DM& LFV\\
         \hline
        A&$y_i \gg y_i'$&Yes&No& No &Yes& Yes&Yes\\
         \hline
B&$y_i \ll y_i'$&Yes&Yes&Not applicable&No& Yes& No\\
\hline        C&$y_i\simeq y_i'$&Yes&No&Yes& No&Yes &No\\
\hline
    \end{tabular}
    \caption{Summary of results for three benchmark scenarios.}
    \label{tab:summary}
\end{centering}
\end{table}

\section{Conclusion}
\label{sec:conclude}
We have proposed a new model of light Dirac neutrinos based on non-Abelian discrete flavour symmetry $A_4$ augmented with $Z_2\times U(1)_L$ where the symmetry breaking pattern 
leaves a remnant $\mathcal{Z}_2 \subset A_4$ symmetry responsible for stabilising DM. Using the minimal flavon and additional fermion content required for realising neutrino data, we take both tree level and one-loop contributions to light Dirac neutrino masses. While tree level or type-I seesaw leads to a rank-1 Dirac neutrino mass matrix with only one non-zero mass eigenvalue, scotogenic or one-loop contributions can generate the other masses. Our proposal, therefore, provides the first generalisation of discrete dark matter and scoto-seesaw scenarios to light Dirac neutrinos.

The remnant $\mathcal{Z}_2$ symmetry allows either fermion or scalar DM candidates. While scalar DM is dominantly produced thermally due to gauge and scalar portal interactions, fermion DM production relies primarily on Yukawa portal interactions. The same Yukawa portal interactions can also lead to thermal or non-thermal production of right handed neutrinos, which contribute to the effective relativistic degrees of freedom $\Delta {\rm N}_{\rm eff}$. To illustrate this, we consider three different benchmark scenarios of such Yukawa couplings and perform a detailed study of neutrino mass, mixing together with DM and $\Delta {\rm N}_{\rm eff}$. We also check the scope of other detection prospects like LFV decays. Table \ref{tab:summary} summarises our results. While one of the sub-classes of Yukawa couplings, namely Case-A, allows fermion singlet DM with observable LFV rates, the corresponding $\Delta {\rm N}_{\rm eff}$ remains negligible. The other two sub-classes allow scalar DM only but with observable $\Delta {\rm N}_{\rm eff}$ at future CMB experiments. The LFV rates, however, remain tiny in these two sub-classes. Due to the rich particle spectrum in terms of several scalar doublets and singlets, the model can also have rich collider phenomenology. On the other hand, spontaneous breaking of discrete symmetries can have interesting consequences in cosmology due to the formation of domain walls. Such walls, which disappear due to soft-breaking terms, can emit stochastic gravitational waves, which can be probed in different ongoing and future experiments. The same soft-breaking terms can also lead to the decay of DM, keeping indirect detection prospects alive. We leave a comprehensive analysis of such additional detection prospects of our model to future works.

\acknowledgments
The work of D.B. is supported by the Science and Engineering Research Board (SERB), Government of India grants MTR/2022/000575 and CRG/2022/000603. The work of B.K. has been supported in part by the Polish National Science Center (NCN) under grant 2020/37/B/ST2/02371, the Freedom of Research, Mobility of Science, Research Excellence Initiative of the University of Silesia in Katowice and EU COST Action (CA21136). S.M. acknowledges the financial support from the National Research Foundation (NRF) grant funded by the Korea government (MEST) NRF-2022R1A2C1005050.

\appendix
\section{$A_{4}$ Multiplication rules}\label{apa}
All finite groups are completely characterized by a set of elements called generators of the group and a set of relations, so that all the elements of the group are given as products of the generators. The group $A_4$ consists of the even permutations of four objects and then contains 4!/2 = 12 elements. The generators are S and T with the relations $S^2 = T^3 = (ST)^3 = \mathcal{I}$, then the elements are $1, S, T, ST, TS, T^2, ST^2, STS, TST, T^2S, TST^2,T^2ST$.

 It has four irreducible representations: three one-dimensional and one three-dimensional which are denoted by $\bf{1}, 
	\bf{1'}, \bf{1''}$ and $\bf{3}$ respectively. 
 Under $A_4$, one-dimensional unitary representations are given as,
\begin{eqnarray*}
    &&1\quad S=1 \quad T=1;\\
&&1'\quad S=1\quad T=\omega;\\
&&1''\quad S=1 \quad T=\omega^2;
\end{eqnarray*}
$\omega=e^{i\frac{2\pi}{3}}$, is cube root of unity.
In the basis, where $S$ is a real diagonal,
\begin{eqnarray}
   S= \begin{pmatrix}
    1&0&0\\0&-1&0\\0&0&-1    
    \end{pmatrix}_{3\times 3} \quad T=\begin{pmatrix}
        0&1&0\\0&0&1\\1&0&0
    \end{pmatrix}_{3\times 3}
\end{eqnarray}
 The multiplication rules of the 
	irreducible representations are given by~\cite{Altarelli:2010gt}
	\begin{equation}\label{a4product}
	\bf{1} \otimes \bf{1} = \bf{1}, 
	\bf{1'}\otimes \bf{1'} = \bf{1''},  \bf{1'} \otimes \bf{1''} = \bf{1} ,  \bf{1''} \otimes \bf{1''} = \bf{1'},  \bf{3} \otimes \bf{3} = \bf{1} + \bf{1'} + \bf{1''} + 
	\bf{3}_{1} +    \bf{3}_{2}
	\end{equation}
	where ${\bf 1}$ and ${\bf 2}$ in the subscript correspond to anti-symmetric and symmetric parts, respectively. Now, if we have two  triplets as $ A = (a_1, a_2, a_3)^T$ and $ B =(b_1, b_2, 
	b_3)^T$ respectively, their direct product can be decomposed into the direct sum mentioned above. The product rule  for this  two triplets  in the $S$ diagonal basis\footnote{Here $S$ is a $3 \times 3$ diagonal generator of $A_4$.} can be written as  \cite{He:2006dk, Memenga:2013vc, Ishimori:2010au}
	\begin{eqnarray}
	(A\times B)_{\bf{1}} &\backsim& a_1b_1+a_2 b_2+a_3b_3,\\
	(A\times B)_{\bf{1'}} &\backsim& a_1 b_1 + \omega^2 a_2 b_2 + \omega a_3 b_3,\\
	(A\times B)_{\bf{1''}} &\backsim& a_1 b_1 + \omega a_2 b_2 + \omega^2 a_3 b_3,\\
	(A\times B)_{\bf{3}_{1}} &\backsim& (a_2b_3+a_3b_2,a_3b_1+a_1b_3, a_1b_2+a_2b_1),\label{eq:3s}\\
	(A\times B)_{\bf{3}_{2}} &\backsim& (a_2b_3-a_3b_2, a_3b_1-a_1b_3, a_1b_2-a_2b_1)\label{eq:3a},
	\end{eqnarray}
	where $\omega$ ($=e^{2i\pi/3}$) is the cube root of unity.

\section{The scalar potential}
\label{appen2}
 
The tree-level scalar potential of our model can be written as
\begin{eqnarray}
 V (H, \eta, \chi)&=&\nonumber   -\mu_h^2(H^\dagger H)-\mu_\eta^2(\eta^\dagger\eta)_1-\mu_\chi^2(\chi \chi)_1 + \lambda_{H}(H^\dagger H)^2+ \lambda_{\eta1} (\eta^\dagger\eta)_1 (\eta^\dagger\eta)_1 + \lambda_{\eta 2}(\eta^\dagger\eta)_{1'}(\eta^\dagger\eta)_{1''} \nonumber \\
 && \nonumber +\lambda_{\eta 3}(\eta^\dagger\eta)_{3_1}(\eta^\dagger\eta)_{3_1}+\lambda_{\eta 4}[(\eta^\dagger\eta)_{3_1}(\eta^\dagger\eta)_{3_2}+{\rm h.c.}]+\lambda_{\eta 5}[(\eta^\dagger\eta)_{3_2}(\eta^\dagger\eta)_{3_2}+{\rm h.c.}] \\ &&\nonumber 
 + \lambda_{\eta H1}(\eta^\dagger\eta)_1(H^\dagger H) + \lambda_{\eta H2}(\eta^\dagger H)(H^\dagger \eta) + \lambda_{\eta H3}[(\eta^\dagger H)(\eta^\dagger H)+{\rm h.c.}] \\&&
 \nonumber  + \lambda_{\eta H4}\big [(\eta^\dagger\eta)_{3_1}H^\dagger\eta + {\rm h.c.}\big]+\lambda_{\eta H5}\big [(\eta^\dagger\eta)_{3_2}H^\dagger\eta + {\rm h.c.}\big]+\lambda_{\chi 1}(\chi \chi)_1 (\chi \chi)_1 \\&&
 \nonumber  +  \lambda_{\chi 2}(\chi \chi)_{1'}(\chi \chi)_{1''} + \lambda_{\chi3}(\chi \chi )_{3_1}(\chi\chi)_{3_1} + \lambda_{\chi H1}(\chi \chi)_1(H^\dagger H) + \lambda_{\eta\chi 1}(\eta^\dagger\eta)_1(\chi \chi)_1 \\&&
 \nonumber +\lambda_{\eta\chi 2} \big[(\eta^\dagger\eta)_{1'}(\chi \chi)_{1''}+{\rm h.c.}\big]+\lambda_{\eta\chi3}\big[(\eta^\dagger\eta)_{1''}(\chi \chi)_{1'}+{\rm h.c.}\big]+\lambda_{\eta\chi4}(\eta^\dagger\eta)_{3_1}(\chi \chi)_{3_1}\nonumber\\
    && +\lambda_{\eta\chi5}\big [(\eta^\dagger\eta)_{3_2}(\chi \chi)_{3_1} +{\rm h.c.} \big ]+ \lambda_{\eta H \chi} \big [ (\chi \chi)_{3_1} (H^\dagger \eta) + {\rm h.c.}\big ],
\end{eqnarray}
where, the product of two triplets contracted into one of the two triplet representations of $A_4$ is indicated by $[...]_{3_{1,2}}$, and the product of two triplets contracted into a singlet is represented by $[...]_{1,1',1''}$. Among the $A_4$ triplet scalars $\eta=(\eta_1,\eta_2,\eta_3)$ and $\chi=(\chi_1,\chi_2,\chi_3)$, only $\eta_1$ and $\chi_1$ are getting VEVs $v_\eta$ and $v_\chi$ respectively. The scalar fields can be can be expressed as

\begin{eqnarray}
H&=&\begin{pmatrix}
    h^\pm\\ (v+h+iA)/\sqrt{2}
\end{pmatrix};
\end{eqnarray} 

\begin{eqnarray}
\eta_1&=&\begin{pmatrix}\eta_1^\pm\\ (v_\eta+h_1+iA_1)/\sqrt{2}
    \end{pmatrix},\quad \eta_{2,3}=\begin{pmatrix}
        \eta_{2,3}^\pm\\ (h_{2,3}+iA_{2,3})/\sqrt{2}
    \end{pmatrix};\\
    \chi_1&=& (v_\chi+x_1)/\sqrt{2}, \quad \chi_{2,3}=x_{2,3}/\sqrt{2}.
\end{eqnarray}
The minimisation conditions for the scalars obtaining non-zero VEVs can be written as
\begin{eqnarray}
 \nonumber       \mu^2_H = &&\lambda_H v^2 + (\lambda_{\eta H1}+\lambda_{\eta H2}+2\lambda_{\eta H3}) \frac{v^2_\eta}{2} + \lambda_{\chi H1} \frac{v^2_\chi}{2}, \\
           \mu^2_\eta =&& (\lambda_{\eta 1} +\lambda_{\eta 2}) v^2_\eta + (\lambda_{\eta H1}+\lambda_{\eta H2}+2\lambda_{\eta H3}) \frac{v^2 }{2} + (\lambda_{\eta \chi1} +2(\lambda_{\eta \chi2}+\lambda_{\eta \chi3})) \frac{v^2_\chi}{2},\\
\nonumber      \mu^2_\chi =&& (\lambda_{\chi1}+\lambda_{\chi2}) v^2_\chi + \lambda_{\chi H1} \frac{v^2}{2}+(\lambda_{\eta \chi1} +2(\lambda_{\eta \chi2}+\lambda_{\eta \chi3})) \frac{v^2_\eta}{2}.
\end{eqnarray}
There are seven neutral scalars $(h, h_1, h_2, h_3, x_1, x_2, x_3)$ out of which four are $\mathcal{Z}_2$-odd while the remaining three are $\mathcal{Z}_2$-even. Similarly, there is one physical pseudo-scalar out of the $\mathcal{Z}_2$-even combination of $(A, A_1)$ while $(A_2, A_3)$ constitute $\mathcal{Z}_2$-odd pseudoscalars. The $\mathcal{Z}_2$-even scalar mass matrix squared in the $(h, h_1, x_1)$ basis can be written as
\begin{eqnarray}
    \mathcal{M}^2_{S+}=\begin{pmatrix}
  M^2_{hh} & M^2_{h h_1} & M^2_{hx_1} \\
  M^2_{h h_1} & M^2_{h_1 h_1} & M^2_{h_1 x_1}\\
  M^2_{h x_1} & M^2_{h_1 x_1} & M^2_{x_1 x_1} \\
    \end{pmatrix}_{3\times3},
    \label{matrix1}
\end{eqnarray}
where,
{\begin{equation}
    M^2_{hh} = 2\lambda_H v^2; \,\, M^2_{h h_1} = (\lambda_{\eta H1}+\lambda_{\eta H2}+2\lambda_{\eta H3}) v_\eta v; \,\, M^2_{h x_1} = \lambda_{\chi H1} v_\chi v; \nonumber 
\end{equation}
\begin{equation}
    M^2_{h_1 h_1} = 2(\lambda_{\eta 1}+ \lambda_{\eta 2}) v^2_\eta; \,\,  M^2_{h_1 x_1} = (\lambda_{\eta \chi1} +2(\lambda_{\eta \chi2}+\lambda_{\eta \chi3}))v_\chi v_\eta; \,\, M^2_{x_1 x_1} = 2(\lambda_{\chi 1}+\lambda_{\chi 2})v^2_\chi. \nonumber
\end{equation}}
This mass squared matrix can be diagonalised by $3\times 3$ orthogonal matrix involving three angles. Similarly, the $\mathcal{Z}_2$-odd scalar mass matrix squared in the $(h_2, h_3, x_2, x_3)$ basis can be written as
\begin{eqnarray}
    \mathcal{M}^2_{S-}=\begin{pmatrix}
  M^2_{h_2h_2} & M^2_{h_2 h_3} & M^2_{h_2x_2} & M^2_{h_2x_3} \\
  M^2_{h_2 h_3} & M^2_{h_3 h_3} & M^2_{h_3 x_2} & M^2_{h_3 x_3}\\
 M^2_{h_2x_2} & M^2_{h_3 x_2} & M^2_{x_2 x_2} & M^2_{x_2 x_3}\\
 M^2_{h_2x_3} & M^2_{h_3 x_3}  & M^2_{x_2 x_3} & M^2_{x_3 x_3} \\
    \end{pmatrix}_{4\times4},
    \label{matrix2}
\end{eqnarray}
where, 
    \begin{eqnarray*}
 M^2_{h_2h_2}=&&M_{h_3h_3}^2=\frac{1}{2} ( (- 3 \lambda_{\eta2} + 4 \lambda_{\eta3}) v^2_\eta - 
   3 (\lambda_{\eta\chi2} + \lambda_{\eta\chi3}) v_\chi^2);\\
    M^2_{x_2x_2}=&&M_{x_3x_3}^2=\frac{1}{2} [-3(\lambda_{\eta\chi2}+\lambda_{\eta\chi3}) v_\eta^2 + (-3 \lambda_{\chi2} + 4\lambda_{\chi3}) v_\chi^2]; \\
    M^2_{h_2h_3}=&&3\lambda_{\eta H4} v v_\eta;\\
    M^2_{h_2x_2}=&&M_{h_3x_3}^2= \lambda_{\eta H\chi} v v_\chi + \lambda_{\eta\chi4} v_\eta v_\chi; \\ M_{h_2x_3}^2=&&M_{h_3x_2}^2=\lambda_{\eta H \chi} v v_\chi;\\ M_{x_2x_3}^2=&&\lambda_{\eta H \chi} v v_\eta.
    \end{eqnarray*}
 This mass squared matrix can similarly be diagonalised by a $4\times 4$ orthogonal matrix involving six angles, in general, leading to physical states. In the scalar DM case, the lightest eigenstate of the $4\times 4$ mass matrix given above becomes the DM candidate. Similar mass matrices can be written for both $\mathcal{Z}_2$-odd and $\mathcal{Z}_2$-even pseudoscalars as well.

The mass matrix squared for pseduo-scalars in the ($A,A_1,A_2,A_3$) basis can be written as
\begin{equation}
   \mathcal{M}_{PS}^2=\begin{pmatrix}
       M^2_{AA}&M^2_{AA_1}&0&0\\M^2_{AA_1}&M^2_{A_1A_1}&0&0\\
       0&0&M^2_{A_2A_2}&M^2_{A_2A_3}\\0&0&M^2_{A_2A_3}&M^2_{A_3A_3}
   \end{pmatrix}_{4\times4},
   \label{matrix3}
\end{equation}
where
\begin{eqnarray*}
M^2_{AA}=&&-2\lambda_{\eta H3} v_\eta^2; \quad M^2_{A_1A_1}=-2\lambda_{\eta H3} v^2;\quad M^2_{AA_1}=2 \lambda_{\eta H3} v v_\eta;\\
M_{A_2A_2}^2=&&M_{A_3A_3}^2=\frac{1}{2} [-4 \lambda_{\eta H3} v^2 -( 3 \lambda_{\eta 2}+ 8\lambda_{\eta 5} )v_\eta^2 - 3( \lambda_{\eta\chi2} + \lambda_{\eta\chi3}) v_\chi^2];\\
M^2_{A_2A_3}=&& \lambda_{\eta H4} v v_\eta.
\end{eqnarray*}

Similarly, for the charged scalars, the mass matrix squared in the ($h^\pm, \eta_1^\pm,\eta_2^\pm,\eta_3^\pm$) basis can be expressed as

\begin{equation}
    \mathcal{M}^2_{CS}=\begin{pmatrix}
        M^2_{h^\pm h^\mp}&M^2_{h^\pm \eta_1^\mp}&0&0\\M^2_{h^\pm\eta_1^\mp}&M^2_{\eta_1^\pm\eta_1^\mp}&0&0\\0&0&M^2_{\eta_2^\pm\eta_2^\mp}&M^2_{\eta_2^\pm\eta_3^{\mp}}\\0&0&M^2_{\eta_2^\pm\eta_3^\mp}&M^2_{\eta_3^\pm\eta_3^{\mp}}
    \end{pmatrix}_{4\times4},
    \label{matrix4}
\end{equation}
\begin{eqnarray*}
    M^2_{h^\pm h^\mp}=&&-(\lambda_{\eta H2} + 2 \lambda_{\eta H3}) \frac{v_\eta^2}{2};\quad M^2_{\eta_1^\pm\eta_1^\mp}=- (\lambda_{\eta H2} + 2 \lambda_{\eta H3}) \frac{v^2}{2};\\
M_{\eta_2^\pm\eta_2^\mp}^2=&&M^2_{\eta_3^\pm\eta_3^\mp}=-(\lambda_{\eta H2} + 2 \lambda_{\eta H3}) \frac{v^2}{2} - 3 \lambda_{\eta2} \frac{v_{\eta}^2}{2} - 
  3 (\lambda_{\eta\chi2} + \lambda_{\eta\chi3}) \frac{v_\chi^2}{2};\\
M^2_{h^\pm\eta_1^\mp}=&&(\lambda_{\eta H2} + 2 \lambda_{\eta H3}) \frac{v v_\eta}{2}; \quad M^2_{\eta_2^\pm\eta_3^\mp}=\lambda_{\eta H4} v v_\eta.
\end{eqnarray*}
Clearly, the $\mathcal{Z}_2$-even $2\times 2$ blocks of $\mathcal{M}^2_{PS}$ and $\mathcal{M}^2_{CS}$ have rank 1 indicating the massless Goldstone modes.

{For illustrative purposes, we use the following benchmark parameters related to the scalar sector to estimate the scalar masses numerically while being in agreement with the loop functions used in neutrino mass analysis: $v=174$ GeV, $v_\eta=174$ GeV, $v_{\chi}=4800$ GeV, $\lambda_H=0.188$, $\lambda_{\eta1}=4.9$, $\lambda_{\eta_2}=4.9$, $\lambda_{\eta3}=0.4$, $\lambda_{\eta5}=0.6$, $\lambda_{\eta H1}=5.3$, $\lambda_{\eta H2}=-0.8$, $\lambda_{\eta H3}=-3$, $\lambda_{\eta H4}=1$, $\lambda_{\eta\chi1}=0.001$, $\lambda_{\eta\chi2}=-0.5$, $\lambda_{\eta\chi3}=0.48$, $\lambda_{\eta\chi 4}=0.1$, $\lambda_{\chi1}=0.1$, $\lambda_{\chi2}=0.5$, $\lambda_{\chi3}=0.8$, $\lambda_{\chi H1}=0.02$, $\lambda_{\eta H\chi }=0.5$. The physical scalar masses for the chosen benchmark are shown in table \ref{bmp1}.}

\begin{table}[h!]
    \centering
   \begin{tabular}{|p{4cm}|p{3cm}|p{3cm}|}
    \hline
         Scalars& Masses& Loop functions  \\
         \hline
         $Z_2$-even neutral scalars&$M_{1+}=125$ GeV  $M_{2+}=775$ GeV $M_{3+}=5258$ GeV& \multirow{4}{10em}{$\mathcal{F}_1=1\times10^{-3} \newline \mathcal{F}_2=6.1\times 10^{-4}$ \newline $\mathcal{F}_3=5.96\times10^{-4}$} \\
              \cline{1-2}
         $Z_2$-odd neutral scalars&$M_{1-}=603$ GeV $M_{2-}=4423$ GeV $M_{3-}=733$ GeV $M_{4-}=4437$ GeV& \\
              \cline{1-2}
         $Z_2$-odd charged scalars&$M_{1c}=785$ GeV $M_{2c}=1169$ GeV $M_{3c}=1154$ GeV& \\
              \cline{1-2}
         $Z_2$-odd pseudo-scalars&$M_{1ps}=738$ GeV 
         $M_{2ps}=866$ GeV $M_{3ps}=876$ GeV& \\
         \hline
    \end{tabular}
    \caption{Numerical estimation for the scalar sector with associated loop functions. We use the notation $M_i$ for the respective mass eigenvalues of scalar mass matrices shown in Eq. \eqref{matrix1}, \eqref{matrix2}, \eqref{matrix3}, \eqref{matrix4}.}
    \label{bmp1}
\end{table}

\section{Neutrino mass matrix}
\label{appen3}
Combining the tree level and one-loop contributions, the complete neutrino mass matrix can be written as
\begin{eqnarray}
m_{\nu}&=&m_{\nu 0}+m^{\rm loop}_{\nu } \\
  &=&\begin{pmatrix}
           (m_{\nu})_{11}  &  (m_{\nu})_{12}  & (m_{\nu})_{13}\\
		(m_{\nu})_{21}&  (m_{\nu})_{22} & (m_{\nu})_{23}\\
		(m_{\nu})_{31} &  (m_{\nu})_{32} & (m_{\nu})_{33}
    \end{pmatrix} ,
\end{eqnarray}
where,
\begin{eqnarray}
(m_{\nu})_{11} &=& -\frac{v_{\eta}v_{\chi}}{2M_N}y_1y'_1 +y_1y'_1M_N(\mathcal{F}_1+\mathcal{F}_2+\mathcal{F}_3) \label{eq:m11};\\
(m_{\nu})_{12} &=& -\frac{v_{\eta}v_{\chi}}{2M_N}y_1y'_2+y_1y'_2M_N(\mathcal{F}_1+\omega\mathcal{F}_2+\omega^2\mathcal{F}_3); \\
(m_{\nu})_{13} &= & -\frac{v_{\eta}v_{\chi}}{2M_N}y_1y'_3 +y_1y'_3M_N(\mathcal{F}_1+\omega^2\mathcal{F}_2+\omega\mathcal{F}_3); \\
(m_{\nu})_{21} &= &  -\frac{v_{\eta}v_{\chi}}{2M_N}y_2y'_1+y_2y'_1M_N(\mathcal{F}_1+\omega\mathcal{F}_2+\omega^2\mathcal{F}_3); \\
(m_{\nu})_{22}  &= &  -\frac{v_{\eta}v_{\chi}}{2M_N}y_2y'_2+y_2y'_2M_N(\mathcal{F}_1+\omega^2\mathcal{F}_2+\omega\mathcal{F}_3); \\
(m_{\nu})_{23} &= &  -\frac{v_{\eta}v_{\chi}}{2M_N}y_2y'_3+y_2y'_3M_N(\mathcal{F}_1+\mathcal{F}_2+\mathcal{F}_3);\\
(m_{\nu})_{31}  &= &  -\frac{v_{\eta}v_{\chi}}{2M_N}y_3y'_1+y_3y'_1M_N(\mathcal{F}_1+\omega^2\mathcal{F}_2+\omega\mathcal{F}_3);\\
(m_{\nu})_{32}  &= &  -\frac{v_{\eta}v_{\chi}}{2M_N}y_3y'_2+y_3y'_2M_N(\mathcal{F}_1+\mathcal{F}_2+\mathcal{F}_3);  \\
(m_{\nu})_{33}  &= &  -\frac{v_{\eta}v_{\chi}}{2M_N}y_3y'_3 +y_3y'_3M_N(\mathcal{F}_1+\omega\mathcal{F}_2+\omega^2\mathcal{F}_3).\label{eq:m33} 
\end{eqnarray}
Now, substituting $y_2=y_3$ and $y'_2=y'_3$ in Eq. (\ref{eq:m11}) - Eq. (\ref{eq:m33}) and assuming all Yukawa couplings to be real, we find the following relations:
\begin{eqnarray}
    & (m_{\nu})_{11}=(m_{\nu})^*_{11}, ~(m_{\nu})_{12}=(m_{\nu})^*_{13},~ (m_{\nu})_{21}=(m_{\nu})^*_{31}, \nonumber \\
   & (m_{\nu})_{23}=(m_{\nu})^*_{32}, (m_{\nu})_{22}=(m_{\nu})^*_{33}.
\end{eqnarray}
These conditions give rise to a $\mu-\tau$ reflection symmetric mass matrix for Dirac neutrinos with maximal atmospheric mixing angle and Dirac CP phase. 


\begin{thebibliography}{100}

\bibitem{ParticleDataGroup:2020ssz}
{\bf Particle Data Group} Collaboration, P.~A. Zyla et~al., {\it {Review of
  Particle Physics}},  {\em PTEP} {\bf 2020} (2020), no.~8 083C01.

\bibitem{Minkowski:1977sc}
P.~Minkowski, {\it {$\mu \to e\gamma$ at a Rate of One Out of $10^{9}$ Muon
  Decays?}},  {\em Phys. Lett.} {\bf B67} (1977) 421--428.

\bibitem{GellMann:1980vs}
M.~Gell-Mann, P.~Ramond, and R.~Slansky, {\it {Complex Spinors and Unified
  Theories}},  {\em Conf. Proc.} {\bf C790927} (1979) 315--321,
  [\href{http://arxiv.org/abs/1306.4669}{{\tt arXiv:1306.4669}}].

\bibitem{Mohapatra:1979ia}
R.~N. Mohapatra and G.~Senjanovic, {\it {Neutrino Mass and Spontaneous Parity
  Violation}},  {\em Phys. Rev. Lett.} {\bf 44} (1980) 912.

\bibitem{Sawada:1979dis}
O.~Sawada and A.~Sugamoto, eds., {\em {Proceedings: Workshop on the Unified
  Theories and the Baryon Number in the Universe}: {Tsukuba, Japan, February
  13-14, 1979}}, (Tsukuba, Japan), Natl.Lab.High Energy Phys., 1979.

\bibitem{Yanagida:1980xy}
T.~Yanagida, {\it {Horizontal Symmetry and Masses of Neutrinos}},  {\em Prog.
  Theor. Phys.} {\bf 64} (1980) 1103.

\bibitem{Schechter:1980gr}
J.~Schechter and J.~W.~F. Valle, {\it {Neutrino Masses in SU(2) x U(1)
  Theories}},  {\em Phys. Rev.} {\bf D22} (1980) 2227.

\bibitem{Mohapatra:1980yp}
R.~N. Mohapatra and G.~Senjanovic, {\it {Neutrino Masses and Mixings in Gauge
  Models with Spontaneous Parity Violation}},  {\em Phys. Rev.} {\bf D23}
  (1981) 165.

\bibitem{Lazarides:1980nt}
G.~Lazarides, Q.~Shafi, and C.~Wetterich, {\it {Proton Lifetime and Fermion
  Masses in an SO(10) Model}},  {\em Nucl. Phys.} {\bf B181} (1981) 287--300.

\bibitem{Wetterich:1981bx}
C.~Wetterich, {\it {Neutrino Masses and the Scale of B-L Violation}},  {\em
  Nucl. Phys.} {\bf B187} (1981) 343--375.

\bibitem{Schechter:1981cv}
J.~Schechter and J.~W.~F. Valle, {\it {Neutrino Decay and Spontaneous Violation
  of Lepton Number}},  {\em Phys. Rev.} {\bf D25} (1982) 774.

\bibitem{Foot:1988aq}
R.~Foot, H.~Lew, X.~G. He, and G.~C. Joshi, {\it {Seesaw Neutrino Masses
  Induced by a Triplet of Leptons}},  {\em Z. Phys.} {\bf C44} (1989) 441.

\bibitem{Roncadelli:1983ty}
M.~Roncadelli and D.~Wyler, {\it {Naturally Light Dirac Neutrinos in Gauge
  Theories}},  {\em Phys. Lett. B} {\bf 133} (1983) 325--329.

\bibitem{Roy:1983be}
P.~Roy and O.~U. Shanker, {\it {Observable Neutrino Dirac Mass and Supergrand
  Unification}},  {\em Phys. Rev. Lett.} {\bf 52} (1984) 713--716. [Erratum:
  Phys.Rev.Lett. 52, 2190 (1984)].

\bibitem{Babu:1988yq}
K.~S. Babu and X.~G. He, {\it {DIRAC NEUTRINO MASSES AS TWO LOOP RADIATIVE
  CORRECTIONS}},  {\em Mod. Phys. Lett.} {\bf A4} (1989) 61.

\bibitem{Peltoniemi:1992ss}
J.~T. Peltoniemi, D.~Tommasini, and J.~W.~F. Valle, {\it {Reconciling dark
  matter and solar neutrinos}},  {\em Phys. Lett.} {\bf B298} (1993) 383--390.

\bibitem{Chulia:2016ngi}
S.~Centelles~Chuliá, E.~Ma, R.~Srivastava, and J.~W.~F. Valle, {\it {Dirac
  Neutrinos and Dark Matter Stability from Lepton Quarticity}},  {\em Phys.
  Lett.} {\bf B767} (2017) 209--213,
  [\href{http://arxiv.org/abs/1606.04543}{{\tt arXiv:1606.04543}}].

\bibitem{Aranda:2013gga}
A.~Aranda, C.~Bonilla, S.~Morisi, E.~Peinado, and J.~W.~F. Valle, {\it {Dirac
  neutrinos from flavor symmetry}},  {\em Phys. Rev.} {\bf D89} (2014), no.~3
  033001, [\href{http://arxiv.org/abs/1307.3553}{{\tt arXiv:1307.3553}}].

\bibitem{Chen:2015jta}
P.~Chen, G.-J. Ding, A.~D. Rojas, C.~A. Vaquera-Araujo, and J.~W.~F. Valle,
  {\it {Warped flavor symmetry predictions for neutrino physics}},  {\em JHEP}
  {\bf 01} (2016) 007, [\href{http://arxiv.org/abs/1509.06683}{{\tt
  arXiv:1509.06683}}].

\bibitem{Ma:2015mjd}
E.~Ma, N.~Pollard, R.~Srivastava, and M.~Zakeri, {\it {Gauge $B-L$ Model with
  Residual $Z_3$ Symmetry}},  {\em Phys. Lett.} {\bf B750} (2015) 135--138,
  [\href{http://arxiv.org/abs/1507.03943}{{\tt arXiv:1507.03943}}].

\bibitem{Reig:2016ewy}
M.~Reig, J.~W.~F. Valle, and C.~A. Vaquera-Araujo, {\it {Realistic
  $\mathrm{SU(3)_c \otimes SU(3)_L \otimes U(1)_X}$ model with a type II Dirac
  neutrino seesaw mechanism}},  {\em Phys. Rev.} {\bf D94} (2016), no.~3
  033012, [\href{http://arxiv.org/abs/1606.08499}{{\tt arXiv:1606.08499}}].

\bibitem{Wang:2016lve}
W.~Wang and Z.-L. Han, {\it {Naturally Small Dirac Neutrino Mass with
  Intermediate $SU(2)_{L}$ Multiplet Fields}},
  \href{http://arxiv.org/abs/1611.03240}{{\tt arXiv:1611.03240}}.
  [JHEP04,166(2017)].

\bibitem{Wang:2017mcy}
W.~Wang, R.~Wang, Z.-L. Han, and J.-Z. Han, {\it {The $B-L$ Scotogenic Models
  for Dirac Neutrino Masses}},  {\em Eur. Phys. J.} {\bf C77} (2017), no.~12
  889, [\href{http://arxiv.org/abs/1705.00414}{{\tt arXiv:1705.00414}}].

\bibitem{Wang:2006jy}
F.~Wang, W.~Wang, and J.~M. Yang, {\it {Split two-Higgs-doublet model and
  neutrino condensation}},  {\em Europhys. Lett.} {\bf 76} (2006) 388--394,
  [\href{http://arxiv.org/abs/hep-ph/0601018}{{\tt hep-ph/0601018}}].

\bibitem{Gabriel:2006ns}
S.~Gabriel and S.~Nandi, {\it {A New two Higgs doublet model}},  {\em Phys.
  Lett.} {\bf B655} (2007) 141--147,
  [\href{http://arxiv.org/abs/hep-ph/0610253}{{\tt hep-ph/0610253}}].

\bibitem{Davidson:2009ha}
S.~M. Davidson and H.~E. Logan, {\it {Dirac neutrinos from a second Higgs
  doublet}},  {\em Phys. Rev.} {\bf D80} (2009) 095008,
  [\href{http://arxiv.org/abs/0906.3335}{{\tt arXiv:0906.3335}}].

\bibitem{Davidson:2010sf}
S.~M. Davidson and H.~E. Logan, {\it {LHC phenomenology of a two-Higgs-doublet
  neutrino mass model}},  {\em Phys. Rev.} {\bf D82} (2010) 115031,
  [\href{http://arxiv.org/abs/1009.4413}{{\tt arXiv:1009.4413}}].

\bibitem{Bonilla:2016zef}
C.~Bonilla and J.~W.~F. Valle, {\it {Naturally light neutrinos in $Diracon$
  model}},  {\em Phys. Lett.} {\bf B762} (2016) 162--165,
  [\href{http://arxiv.org/abs/1605.08362}{{\tt arXiv:1605.08362}}].

\bibitem{Farzan:2012sa}
Y.~Farzan and E.~Ma, {\it {Dirac neutrino mass generation from dark matter}},
  {\em Phys. Rev.} {\bf D86} (2012) 033007,
  [\href{http://arxiv.org/abs/1204.4890}{{\tt arXiv:1204.4890}}].

\bibitem{Bonilla:2016diq}
C.~Bonilla, E.~Ma, E.~Peinado, and J.~W.~F. Valle, {\it {Two-loop Dirac
  neutrino mass and WIMP dark matter}},  {\em Phys. Lett.} {\bf B762} (2016)
  214--218, [\href{http://arxiv.org/abs/1607.03931}{{\tt arXiv:1607.03931}}].

\bibitem{Ma:2016mwh}
E.~Ma and O.~Popov, {\it {Pathways to Naturally Small Dirac Neutrino Masses}},
  {\em Phys. Lett.} {\bf B764} (2017) 142--144,
  [\href{http://arxiv.org/abs/1609.02538}{{\tt arXiv:1609.02538}}].

\bibitem{Ma:2017kgb}
E.~Ma and U.~Sarkar, {\it {Radiative Left-Right Dirac Neutrino Mass}},  {\em
  Phys. Lett.} {\bf B776} (2018) 54--57,
  [\href{http://arxiv.org/abs/1707.07698}{{\tt arXiv:1707.07698}}].

\bibitem{Borah:2016lrl}
D.~Borah, {\it {Light sterile neutrino and dark matter in left-right symmetric
  models without a Higgs bidoublet}},  {\em Phys. Rev.} {\bf D94} (2016), no.~7
  075024, [\href{http://arxiv.org/abs/1607.00244}{{\tt arXiv:1607.00244}}].

\bibitem{Borah:2016zbd}
D.~Borah and A.~Dasgupta, {\it {Common Origin of Neutrino Mass, Dark Matter and
  Dirac Leptogenesis}},  {\em JCAP} {\bf 1612} (2016), no.~12 034,
  [\href{http://arxiv.org/abs/1608.03872}{{\tt arXiv:1608.03872}}].

\bibitem{Borah:2016hqn}
D.~Borah and A.~Dasgupta, {\it {Observable Lepton Number Violation with
  Predominantly Dirac Nature of Active Neutrinos}},  {\em JHEP} {\bf 01} (2017)
  072, [\href{http://arxiv.org/abs/1609.04236}{{\tt arXiv:1609.04236}}].

\bibitem{Borah:2017leo}
D.~Borah and A.~Dasgupta, {\it {Naturally Light Dirac Neutrino in Left-Right
  Symmetric Model}},  {\em JCAP} {\bf 1706} (2017), no.~06 003,
  [\href{http://arxiv.org/abs/1702.02877}{{\tt arXiv:1702.02877}}].

\bibitem{CentellesChulia:2017koy}
S.~Centelles~Chuliá, R.~Srivastava, and J.~W.~F. Valle, {\it {Generalized
  Bottom-Tau unification, neutrino oscillations and dark matter: predictions
  from a lepton quarticity flavor approach}},  {\em Phys. Lett.} {\bf B773}
  (2017) 26--33, [\href{http://arxiv.org/abs/1706.00210}{{\tt
  arXiv:1706.00210}}].

\bibitem{Bonilla:2017ekt}
C.~Bonilla, J.~M. Lamprea, E.~Peinado, and J.~W.~F. Valle, {\it
  {Flavour-symmetric type-II Dirac neutrino seesaw mechanism}},  {\em Phys.
  Lett.} {\bf B779} (2018) 257--261,
  [\href{http://arxiv.org/abs/1710.06498}{{\tt arXiv:1710.06498}}].

\bibitem{Memenga:2013vc}
N.~Memenga, W.~Rodejohann, and H.~Zhang, {\it {$A_4$ flavor symmetry model for
  Dirac neutrinos and sizable $U_{e3}$}},  {\em Phys. Rev.} {\bf D87} (2013),
  no.~5 053021, [\href{http://arxiv.org/abs/1301.2963}{{\tt arXiv:1301.2963}}].

\bibitem{Borah:2017dmk}
D.~Borah and B.~Karmakar, {\it {$A_4$ flavour model for Dirac neutrinos: Type I
  and inverse seesaw}},  {\em Phys. Lett.} {\bf B780} (2018) 461--470,
  [\href{http://arxiv.org/abs/1712.06407}{{\tt arXiv:1712.06407}}].

\bibitem{CentellesChulia:2018gwr}
S.~Centelles~Chuliá, R.~Srivastava, and J.~W.~F. Valle, {\it {Seesaw roadmap
  to neutrino mass and dark matter}},  {\em Phys. Lett.} {\bf B781} (2018)
  122--128, [\href{http://arxiv.org/abs/1802.05722}{{\tt arXiv:1802.05722}}].

\bibitem{CentellesChulia:2018bkz}
S.~Centelles~Chuliá, R.~Srivastava, and J.~W.~F. Valle, {\it {Seesaw Dirac
  neutrino mass through dimension-6 operators}},
  \href{http://arxiv.org/abs/1804.03181}{{\tt arXiv:1804.03181}}.

\bibitem{Han:2018zcn}
Z.-L. Han and W.~Wang, {\it {$Z'$ Portal Dark Matter in $B-L$ Scotogenic Dirac
  Model}},  \href{http://arxiv.org/abs/1805.02025}{{\tt arXiv:1805.02025}}.

\bibitem{Borah:2018gjk}
D.~Borah, B.~Karmakar, and D.~Nanda, {\it {Common Origin of Dirac Neutrino Mass
  and Freeze-in Massive Particle Dark Matter}},  {\em JCAP} {\bf 1807} (2018),
  no.~07 039, [\href{http://arxiv.org/abs/1805.11115}{{\tt arXiv:1805.11115}}].

\bibitem{Borah:2018nvu}
D.~Borah and B.~Karmakar, {\it {Linear seesaw for Dirac neutrinos with $A_4$
  flavour symmetry}},  {\em Phys. Lett.} {\bf B789} (2019) 59--70,
  [\href{http://arxiv.org/abs/1806.10685}{{\tt arXiv:1806.10685}}].

\bibitem{Borah:2019bdi}
D.~Borah, D.~Nanda, and A.~K. Saha, {\it {Common origin of modified chaotic
  inflation, non thermal dark matter and Dirac neutrino mass}},
  \href{http://arxiv.org/abs/1904.04840}{{\tt arXiv:1904.04840}}.

\bibitem{CentellesChulia:2019xky}
S.~Centelles~Chuliá, R.~Cepedello, E.~Peinado, and R.~Srivastava, {\it
  {Systematic classification of two loop $d$ = 4 Dirac neutrino mass models and
  the Diracness-dark matter stability connection}},  {\em JHEP} {\bf 10} (2019)
  093, [\href{http://arxiv.org/abs/1907.08630}{{\tt arXiv:1907.08630}}].

\bibitem{Jana:2019mgj}
S.~Jana, V.~P. K., and S.~Saad, {\it {Minimal Realizations of Dirac Neutrino
  Mass from Generic One-loop and Two-loop Topologies at $d=5$}},
  \href{http://arxiv.org/abs/1910.09537}{{\tt arXiv:1910.09537}}.

\bibitem{Nanda:2019nqy}
D.~Nanda and D.~Borah, {\it {Connecting Light Dirac Neutrinos to a
  Multi-component Dark Matter Scenario in Gauged $B-L$ Model}},
  \href{http://arxiv.org/abs/1911.04703}{{\tt arXiv:1911.04703}}.

\bibitem{Guo:2020qin}
S.-Y. Guo and Z.-L. Han, {\it {Observable Signatures of Scotogenic Dirac
  Model}},  {\em JHEP} {\bf 12} (2020) 062,
  [\href{http://arxiv.org/abs/2005.08287}{{\tt arXiv:2005.08287}}].

\bibitem{Bernal:2021ezl}
N.~Bernal, J.~Calle, and D.~Restrepo, {\it {Anomaly-free Abelian gauge
  symmetries with Dirac scotogenic models}},  {\em Phys. Rev. D} {\bf 103}
  (2021), no.~9 095032, [\href{http://arxiv.org/abs/2102.06211}{{\tt
  arXiv:2102.06211}}].

\bibitem{Borah:2022obi}
D.~Borah, S.~Mahapatra, D.~Nanda, and N.~Sahu, {\it {Type II Dirac seesaw with
  observable \ensuremath{\Delta}Neff in the light of W-mass anomaly}},  {\em
  Phys. Lett. B} {\bf 833} (2022) 137297,
  [\href{http://arxiv.org/abs/2204.08266}{{\tt arXiv:2204.08266}}].

\bibitem{Li:2022chc}
S.-P. Li, X.-Q. Li, X.-S. Yan, and Y.-D. Yang, {\it {Scotogenic Dirac neutrino
  mass models embedded with leptoquarks: one pathway to address the flavor
  anomalies and the neutrino masses together}},  {\em Eur. Phys. J. C} {\bf 82}
  (2022), no.~11 1078, [\href{http://arxiv.org/abs/2204.09201}{{\tt
  arXiv:2204.09201}}].

\bibitem{Dey:2024ctx}
M.~Dey and S.~Roy, {\it {Revisiting the Dirac Nature of Neutrinos}},
  \href{http://arxiv.org/abs/2403.12461}{{\tt arXiv:2403.12461}}.

\bibitem{Singh:2024imk}
L.~Singh, M.~Kashav, and S.~Verma, {\it {Minimal Type-I Dirac seesaw and
  Leptogenesis under $A_{4}$ modular invariance}},
  \href{http://arxiv.org/abs/2405.07165}{{\tt arXiv:2405.07165}}.

\bibitem{Zwicky:1933gu}
F.~Zwicky, {\it {Die Rotverschiebung von extragalaktischen Nebeln}},  {\em
  Helv. Phys. Acta} {\bf 6} (1933) 110--127. [Gen. Rel. Grav.41,207(2009)].

\bibitem{Rubin:1970zza}
V.~C. Rubin and W.~K. Ford, Jr., {\it {Rotation of the Andromeda Nebula from a
  Spectroscopic Survey of Emission Regions}},  {\em Astrophys. J.} {\bf 159}
  (1970) 379--403.

\bibitem{Clowe:2006eq}
D.~Clowe, M.~Bradac, A.~H. Gonzalez, M.~Markevitch, S.~W. Randall, C.~Jones,
  and D.~Zaritsky, {\it {A direct empirical proof of the existence of dark
  matter}},  {\em Astrophys. J.} {\bf 648} (2006) L109--L113,
  [\href{http://arxiv.org/abs/astro-ph/0608407}{{\tt astro-ph/0608407}}].

\bibitem{Planck:2018vyg}
{\bf Planck} Collaboration, N.~Aghanim et~al., {\it {Planck 2018 results. VI.
  Cosmological parameters}},  {\em Astron. Astrophys.} {\bf 641} (2020) A6,
  [\href{http://arxiv.org/abs/1807.06209}{{\tt arXiv:1807.06209}}]. [Erratum:
  Astron.Astrophys. 652, C4 (2021)].

\bibitem{LUX-ZEPLIN:2022qhg}
{\bf LUX-ZEPLIN} Collaboration, J.~Aalbers et~al., {\it {First Dark Matter
  Search Results from the LUX-ZEPLIN (LZ) Experiment}},
  \href{http://arxiv.org/abs/2207.03764}{{\tt arXiv:2207.03764}}.

\bibitem{Arcadi:2017kky}
G.~Arcadi, M.~Dutra, P.~Ghosh, M.~Lindner, Y.~Mambrini, M.~Pierre, S.~Profumo,
  and F.~S. Queiroz, {\it {The Waning of the WIMP? A Review of Models,
  Searches, and Constraints}},  \href{http://arxiv.org/abs/1703.07364}{{\tt
  arXiv:1703.07364}}.

\bibitem{Bernal:2017kxu}
N.~Bernal, M.~Heikinheimo, T.~Tenkanen, K.~Tuominen, and V.~Vaskonen, {\it {The
  Dawn of FIMP Dark Matter: A Review of Models and Constraints}},  {\em Int. J.
  Mod. Phys.} {\bf A32} (2017), no.~27 1730023,
  [\href{http://arxiv.org/abs/1706.07442}{{\tt arXiv:1706.07442}}].

\bibitem{Altarelli:2010gt}
G.~Altarelli and F.~Feruglio, {\it {Discrete Flavor Symmetries and Models of
  Neutrino Mixing}},  {\em Rev. Mod. Phys.} {\bf 82} (2010) 2701--2729,
  [\href{http://arxiv.org/abs/1002.0211}{{\tt arXiv:1002.0211}}].

\bibitem{King:2013eh}
S.~F. King and C.~Luhn, {\it {Neutrino Mass and Mixing with Discrete
  Symmetry}},  {\em Rept. Prog. Phys.} {\bf 76} (2013) 056201,
  [\href{http://arxiv.org/abs/1301.1340}{{\tt arXiv:1301.1340}}].

\bibitem{Petcov:2017ggy}
S.~T. Petcov, {\it {Discrete Flavour Symmetries, Neutrino Mixing and Leptonic
  CP Violation}},  {\em Eur. Phys. J. C} {\bf 78} (2018), no.~9 709,
  [\href{http://arxiv.org/abs/1711.10806}{{\tt arXiv:1711.10806}}].

\bibitem{Chauhan:2023faf}
G.~Chauhan, P.~S.~B. Dev, I.~Dubovyk, B.~Dziewit, W.~Flieger, K.~Grzanka,
  J.~Gluza, B.~Karmakar, and S.~Zieba, {\it {Phenomenology of Lepton Masses and
  Mixing with Discrete Flavor Symmetries}},
  \href{http://arxiv.org/abs/2310.20681}{{\tt arXiv:2310.20681}}.

\bibitem{Ding:2024ozt}
G.-J. Ding and J.~W.~F. Valle, {\it {The symmetry approach to quark and lepton
  masses and mixing}},  \href{http://arxiv.org/abs/2402.16963}{{\tt
  arXiv:2402.16963}}.

\bibitem{Hirsch:2010ru}
M.~Hirsch, S.~Morisi, E.~Peinado, and J.~W.~F. Valle, {\it {Discrete dark
  matter}},  {\em Phys. Rev. D} {\bf 82} (2010) 116003,
  [\href{http://arxiv.org/abs/1007.0871}{{\tt arXiv:1007.0871}}].

\bibitem{Meloni:2010sk}
D.~Meloni, S.~Morisi, and E.~Peinado, {\it {Neutrino phenomenology and stable
  dark matter with A4}},  {\em Phys. Lett. B} {\bf 697} (2011) 339--342,
  [\href{http://arxiv.org/abs/1011.1371}{{\tt arXiv:1011.1371}}].

\bibitem{Boucenna:2011tj}
M.~S. Boucenna, M.~Hirsch, S.~Morisi, E.~Peinado, M.~Taoso, and J.~W.~F. Valle,
  {\it {Phenomenology of Dark Matter from $A_4$ Flavor Symmetry}},  {\em JHEP}
  {\bf 05} (2011) 037, [\href{http://arxiv.org/abs/1101.2874}{{\tt
  arXiv:1101.2874}}].

\bibitem{Adulpravitchai:2011ei}
A.~Adulpravitchai, B.~Batell, and J.~Pradler, {\it {Non-Abelian Discrete Dark
  Matter}},  {\em Phys. Lett. B} {\bf 700} (2011) 207--216,
  [\href{http://arxiv.org/abs/1103.3053}{{\tt arXiv:1103.3053}}].

\bibitem{Eby:2011qa}
D.~A. Eby and P.~H. Frampton, {\it {Dark Matter from Binary Tetrahedral Flavor
  Symmetry}},  {\em Phys. Lett. B} {\bf 713} (2012) 249--254,
  [\href{http://arxiv.org/abs/1111.4938}{{\tt arXiv:1111.4938}}].

\bibitem{Boucenna:2012qb}
M.~S. Boucenna, S.~Morisi, E.~Peinado, Y.~Shimizu, and J.~W.~F. Valle, {\it
  {Predictive discrete dark matter model and neutrino oscillations}},  {\em
  Phys. Rev. D} {\bf 86} (2012) 073008,
  [\href{http://arxiv.org/abs/1204.4733}{{\tt arXiv:1204.4733}}].

\bibitem{Hamada:2014xha}
Y.~Hamada, T.~Kobayashi, A.~Ogasahara, Y.~Omura, F.~Takayama, and D.~Yasuhara,
  {\it {Revisiting discrete dark matter model: $\theta_{13} \neq 0$ and
  $\nu_{R}$ dark matter}},  {\em JHEP} {\bf 10} (2014) 183,
  [\href{http://arxiv.org/abs/1405.3592}{{\tt arXiv:1405.3592}}].

\bibitem{Lamprea:2016egz}
J.~M. Lamprea and E.~Peinado, {\it {Seesaw scale discrete dark matter and
  two-zero texture Majorana neutrino mass matrices}},  {\em Phys. Rev. D} {\bf
  94} (2016), no.~5 055007, [\href{http://arxiv.org/abs/1603.02190}{{\tt
  arXiv:1603.02190}}].

\bibitem{DeLaVega:2018bkp}
L.~M.~G. De~La~Vega, R.~Ferro-Hernandez, and E.~Peinado, {\it {Simple $A_4$
  models for dark matter stability with texture zeros}},  {\em Phys. Rev. D}
  {\bf 99} (2019), no.~5 055044, [\href{http://arxiv.org/abs/1811.10619}{{\tt
  arXiv:1811.10619}}].

\bibitem{Bonilla:2023pna}
C.~Bonilla, J.~Herms, O.~Medina, and E.~Peinado, {\it {Discrete dark matter
  mechanism as the source of neutrino mass scales}},  {\em JHEP} {\bf 06}
  (2023) 078, [\href{http://arxiv.org/abs/2301.10811}{{\tt arXiv:2301.10811}}].

\bibitem{Kumar:2024zfb}
R.~Kumar, N.~Nath, and R.~Srivastava, {\it {Cutting the Scotogenic loop: Adding
  flavor to Dark Matter}},  \href{http://arxiv.org/abs/2406.00188}{{\tt
  arXiv:2406.00188}}.

\bibitem{Rojas:2018wym}
N.~Rojas, R.~Srivastava, and J.~W.~F. Valle, {\it {Simplest Scoto-Seesaw
  Mechanism}},  {\em Phys. Lett. B} {\bf 789} (2019) 132--136,
  [\href{http://arxiv.org/abs/1807.11447}{{\tt arXiv:1807.11447}}].

\bibitem{Abazajian:2019eic}
K.~Abazajian et~al., {\it {CMB-S4 Science Case, Reference Design, and Project
  Plan}},  \href{http://arxiv.org/abs/1907.04473}{{\tt arXiv:1907.04473}}.

\bibitem{SPT-3G:2014dbx}
{\bf SPT-3G} Collaboration, B.~A. Benson et~al., {\it {SPT-3G: A
  Next-Generation Cosmic Microwave Background Polarization Experiment on the
  South Pole Telescope}},  {\em Proc. SPIE Int. Soc. Opt. Eng.} {\bf 9153}
  (2014) 91531P, [\href{http://arxiv.org/abs/1407.2973}{{\tt
  arXiv:1407.2973}}].

\bibitem{SimonsObservatory:2018koc}
{\bf Simons Observatory} Collaboration, P.~Ade et~al., {\it {The Simons
  Observatory: Science goals and forecasts}},  {\em JCAP} {\bf 02} (2019) 056,
  [\href{http://arxiv.org/abs/1808.07445}{{\tt arXiv:1808.07445}}].

\bibitem{CMB-HD:2022bsz}
{\bf CMB-HD} Collaboration, S.~Aiola et~al., {\it {Snowmass2021 CMB-HD White
  Paper}},  \href{http://arxiv.org/abs/2203.05728}{{\tt arXiv:2203.05728}}.

\bibitem{CentellesChulia:2016rms}
S.~Centelles~Chuli\'a, E.~Ma, R.~Srivastava, and J.~W.~F. Valle, {\it {Dirac
  Neutrinos and Dark Matter Stability from Lepton Quarticity}},  {\em Phys.
  Lett. B} {\bf 767} (2017) 209--213,
  [\href{http://arxiv.org/abs/1606.04543}{{\tt arXiv:1606.04543}}].

\bibitem{Tao:1996vb}
Z.-j. Tao, {\it {Radiative seesaw mechanism at weak scale}},  {\em Phys. Rev.
  D} {\bf 54} (1996) 5693--5697,
  [\href{http://arxiv.org/abs/hep-ph/9603309}{{\tt hep-ph/9603309}}].

\bibitem{Ma:2006fn}
E.~Ma, {\it {Common origin of neutrino mass, dark matter, and baryogenesis}},
  {\em Mod. Phys. Lett.} {\bf A21} (2006) 1777--1782,
  [\href{http://arxiv.org/abs/hep-ph/0605180}{{\tt hep-ph/0605180}}].

\bibitem{Gu:2007ug}
P.-H. Gu and U.~Sarkar, {\it {Radiative Neutrino Mass, Dark Matter and
  Leptogenesis}},  {\em Phys. Rev. D} {\bf 77} (2008) 105031,
  [\href{http://arxiv.org/abs/0712.2933}{{\tt arXiv:0712.2933}}].

\bibitem{Ma:2019iwj}
E.~Ma, {\it {Scotogenic Cobimaximal Dirac Neutrino Mixing from $\Delta(27)$ and
  $U(1)_\chi$}},  \href{http://arxiv.org/abs/1905.01535}{{\tt
  arXiv:1905.01535}}.

\bibitem{Ma:2019yfo}
E.~Ma, {\it {Scotogenic $U(1)_\chi$ Dirac neutrinos}},  {\em Phys. Lett. B}
  {\bf 793} (2019) 411--414, [\href{http://arxiv.org/abs/1901.09091}{{\tt
  arXiv:1901.09091}}].

\bibitem{Ma:2019coj}
E.~Ma, {\it {Leptonic Source of Dark Matter and Radiative Majorana or Dirac
  Neutrino Mass}},  {\em Phys. Lett. B} {\bf 809} (2020) 135736,
  [\href{http://arxiv.org/abs/1912.11950}{{\tt arXiv:1912.11950}}].

\bibitem{Leite:2020wjl}
J.~Leite, A.~Morales, J.~W.~F. Valle, and C.~A. Vaquera-Araujo, {\it
  {Scotogenic dark matter and Dirac neutrinos from unbroken gauged B
  \ensuremath{-} L symmetry}},  {\em Phys. Lett. B} {\bf 807} (2020) 135537,
  [\href{http://arxiv.org/abs/2003.02950}{{\tt arXiv:2003.02950}}].

\bibitem{Chowdhury:2022jde}
T.~A. Chowdhury, M.~Ehsanuzzaman, and S.~Saad, {\it {Dark Matter and $(g-2)_{\mu, e}$ in radiative Dirac neutrino mass models}},  {\em JCAP} {\bf 08}
  (2022) 076, [\href{http://arxiv.org/abs/2203.14983}{{\tt arXiv:2203.14983}}].

\bibitem{Borah:2022phw}
D.~Borah, E.~Ma, and D.~Nanda, {\it {Dark SU(2) gauge symmetry and scotogenic
  Dirac neutrinos}},  {\em Phys. Lett. B} {\bf 835} (2022) 137539,
  [\href{http://arxiv.org/abs/2204.13205}{{\tt arXiv:2204.13205}}].

\bibitem{Borah:2022enh}
D.~Borah, P.~Das, and D.~Nanda, {\it {Observable $\Delta
  {\textrm{N}}_{\textrm{eff}}$ in Dirac scotogenic model}},  {\em Eur. Phys. J.
  C} {\bf 84} (2024), no.~2 140, [\href{http://arxiv.org/abs/2211.13168}{{\tt
  arXiv:2211.13168}}].

\bibitem{Harrison:2002et}
P.~F. Harrison and W.~G. Scott, {\it {mu - tau reflection symmetry in lepton
  mixing and neutrino oscillations}},  {\em Phys. Lett. B} {\bf 547} (2002)
  219--228, [\href{http://arxiv.org/abs/hep-ph/0210197}{{\tt hep-ph/0210197}}].

\bibitem{Mandal:2021yph}
S.~Mandal, R.~Srivastava, and J.~W.~F. Valle, {\it {The simplest scoto-seesaw
  model: WIMP dark matter phenomenology and Higgs vacuum stability}},  {\em
  Phys. Lett. B} {\bf 819} (2021) 136458,
  [\href{http://arxiv.org/abs/2104.13401}{{\tt arXiv:2104.13401}}].

\bibitem{Barreiros:2020gxu}
D.~M. Barreiros, F.~R. Joaquim, R.~Srivastava, and J.~W.~F. Valle, {\it
  {Minimal scoto-seesaw mechanism with spontaneous CP violation}},  {\em JHEP}
  {\bf 04} (2021) 249, [\href{http://arxiv.org/abs/2012.05189}{{\tt
  arXiv:2012.05189}}].

\bibitem{Barreiros:2022aqu}
D.~M. Barreiros, H.~B. Camara, and F.~R. Joaquim, {\it {Flavour and dark matter
  in a scoto/type-II seesaw model}},  {\em JHEP} {\bf 08} (2022) 030,
  [\href{http://arxiv.org/abs/2204.13605}{{\tt arXiv:2204.13605}}].

\bibitem{Ganguly:2022qxj}
J.~Ganguly, J.~Gluza, and B.~Karmakar, {\it {Common origin of
  \ensuremath{\theta}$_{13}$ and dark matter within the flavor symmetric
  scoto-seesaw framework}},  {\em JHEP} {\bf 11} (2022) 074,
  [\href{http://arxiv.org/abs/2209.08610}{{\tt arXiv:2209.08610}}].

\bibitem{Ganguly:2023jml}
J.~Ganguly, J.~Gluza, B.~Karmakar, and S.~Mahapatra, {\it {Phenomenology of the
  flavor symmetric scoto-seesaw model with dark matter and TM$_1$ mixing}},
  \href{http://arxiv.org/abs/2311.15997}{{\tt arXiv:2311.15997}}.

\bibitem{Kumar:2023moh}
R.~Kumar, P.~Mishra, M.~K. Behera, R.~Mohanta, and R.~Srivastava, {\it
  {Predictions from scoto-seesaw with A4 modular symmetry}},  {\em Phys. Lett.
  B} {\bf 853} (2024) 138635, [\href{http://arxiv.org/abs/2310.02363}{{\tt
  arXiv:2310.02363}}].

\bibitem{Esteban:2020cvm}
I.~Esteban, M.~Gonzalez-Garcia, M.~Maltoni, T.~Schwetz, and A.~Zhou, {\it {The
  fate of hints: updated global analysis of three-flavor neutrino
  oscillations}},  {\em JHEP} {\bf 09} (2020) 178,
  [\href{http://arxiv.org/abs/2007.14792}{{\tt arXiv:2007.14792}}].

\bibitem{deSalas:2020pgw}
P.~F. de~Salas, D.~V. Forero, S.~Gariazzo, P.~Mart\'\i{}nez-Mirav\'e, O.~Mena,
  C.~A. Ternes, M.~T\'ortola, and J.~W.~F. Valle, {\it {2020 global
  reassessment of the neutrino oscillation picture}},  {\em JHEP} {\bf 02}
  (2021) 071, [\href{http://arxiv.org/abs/2006.11237}{{\tt arXiv:2006.11237}}].

\bibitem{Ahriche:2017iar}
A.~Ahriche, A.~Jueid, and S.~Nasri, {\it {Radiative neutrino mass and Majorana
  dark matter within an inert Higgs doublet model}},  {\em Phys. Rev.} {\bf
  D97} (2018), no.~9 095012, [\href{http://arxiv.org/abs/1710.03824}{{\tt
  arXiv:1710.03824}}].

\bibitem{Borah:2018smz}
D.~Borah, D.~Nanda, N.~Narendra, and N.~Sahu, {\it {Right-handed neutrino dark
  matter with radiative neutrino mass in gauged B-L model}},
   {\em Nucl. Phys. B} {\bf 950} (2020) 114841,
  [\href{http://arxiv.org/abs/1810.12920}{{\tt arXiv:1810.12920}}].

\bibitem{Mahanta:2019gfe}
D.~Mahanta and D.~Borah, {\it {Fermion dark matter with $N_2$ leptogenesis in
  minimal scotogenic model}},  {\em JCAP} {\bf 11} (2019) 021,
  [\href{http://arxiv.org/abs/1906.03577}{{\tt arXiv:1906.03577}}].

\bibitem{Alguero:2023zol}
G.~Alguero, G.~Belanger, F.~Boudjema, S.~Chakraborti, A.~Goudelis, S.~Kraml,
  A.~Mjallal, and A.~Pukhov, {\it {micrOMEGAs 6.0: N-component dark matter}},
  {\em Comput. Phys. Commun.} {\bf 299} (2024) 109133,
  [\href{http://arxiv.org/abs/2312.14894}{{\tt arXiv:2312.14894}}].

\bibitem{Semenov:2014rea}
A.~Semenov, {\it {LanHEP — A package for automatic generation of Feynman
  rules from the Lagrangian. Version 3.2}},  {\em Comput. Phys. Commun.} {\bf
  201} (2016) 167--170, [\href{http://arxiv.org/abs/1412.5016}{{\tt
  arXiv:1412.5016}}].

\bibitem{CRESST:2019jnq}
{\bf CRESST} Collaboration, A.~H. Abdelhameed et~al., {\it {First results from
  the CRESST-III low-mass dark matter program}},  {\em Phys. Rev. D} {\bf 100}
  (2019), no.~10 102002, [\href{http://arxiv.org/abs/1904.00498}{{\tt
  arXiv:1904.00498}}].

\bibitem{DarkSide:2018bpj}
{\bf DarkSide} Collaboration, P.~Agnes et~al., {\it {Low-Mass Dark Matter
  Search with the DarkSide-50 Experiment}},  {\em Phys. Rev. Lett.} {\bf 121}
  (2018), no.~8 081307, [\href{http://arxiv.org/abs/1802.06994}{{\tt
  arXiv:1802.06994}}].

\bibitem{DARWIN:2016hyl}
{\bf DARWIN} Collaboration, J.~Aalbers et~al., {\it {DARWIN: towards the
  ultimate dark matter detector}},  {\em JCAP} {\bf 11} (2016) 017,
  [\href{http://arxiv.org/abs/1606.07001}{{\tt arXiv:1606.07001}}].

\bibitem{Griest:1989wd}
K.~Griest and M.~Kamionkowski, {\it {Unitarity Limits on the Mass and Radius of
  Dark Matter Particles}},  {\em Phys. Rev. Lett.} {\bf 64} (1990) 615.

\bibitem{Bhatia:2020itt}
D.~Bhatia and S.~Mukhopadhyay, {\it {Unitarity limits on thermal dark matter in
  (non-)standard cosmologies}},  {\em JHEP} {\bf 03} (2021) 133,
  [\href{http://arxiv.org/abs/2010.09762}{{\tt arXiv:2010.09762}}].

\bibitem{Hall:2009bx}
L.~J. Hall, K.~Jedamzik, J.~March-Russell, and S.~M. West, {\it {Freeze-In
  Production of FIMP Dark Matter}},  {\em JHEP} {\bf 03} (2010) 080,
  [\href{http://arxiv.org/abs/0911.1120}{{\tt arXiv:0911.1120}}].

\bibitem{Cyburt:2015mya}
R.~H. Cyburt, B.~D. Fields, K.~A. Olive, and T.-H. Yeh, {\it {Big Bang
  Nucleosynthesis: 2015}},  {\em Rev. Mod. Phys.} {\bf 88} (2016) 015004,
  [\href{http://arxiv.org/abs/1505.01076}{{\tt arXiv:1505.01076}}].

\bibitem{DESI:2024mwx}
{\bf DESI} Collaboration, A.~G. Adame et~al., {\it {DESI 2024 VI: Cosmological
  Constraints from the Measurements of Baryon Acoustic Oscillations}},
  \href{http://arxiv.org/abs/2404.03002}{{\tt arXiv:2404.03002}}.

\bibitem{Mangano:2005cc}
G.~Mangano, G.~Miele, S.~Pastor, T.~Pinto, O.~Pisanti, and P.~D. Serpico, {\it
  {Relic neutrino decoupling including flavor oscillations}},  {\em Nucl. Phys.
  B} {\bf 729} (2005) 221--234,
  [\href{http://arxiv.org/abs/hep-ph/0506164}{{\tt hep-ph/0506164}}].

\bibitem{Grohs:2015tfy}
E.~Grohs, G.~M. Fuller, C.~T. Kishimoto, M.~W. Paris, and A.~Vlasenko, {\it
  {Neutrino energy transport in weak decoupling and big bang nucleosynthesis}},
   {\em Phys. Rev. D} {\bf 93} (2016), no.~8 083522,
  [\href{http://arxiv.org/abs/1512.02205}{{\tt arXiv:1512.02205}}].

\bibitem{deSalas:2016ztq}
P.~F. de~Salas and S.~Pastor, {\it {Relic neutrino decoupling with flavour
  oscillations revisited}},  {\em JCAP} {\bf 1607} (2016), no.~07 051,
  [\href{http://arxiv.org/abs/1606.06986}{{\tt arXiv:1606.06986}}].

\bibitem{Abazajian:2019oqj}
K.~N. Abazajian and J.~Heeck, {\it {Observing Dirac neutrinos in the cosmic
  microwave background}},  {\em Phys. Rev.} {\bf D100} (2019) 075027,
  [\href{http://arxiv.org/abs/1908.03286}{{\tt arXiv:1908.03286}}].

\bibitem{FileviezPerez:2019cyn}
P.~Fileviez~Pérez, C.~Murgui, and A.~D. Plascencia, {\it {Neutrino-Dark Matter
  Connections in Gauge Theories}},  {\em Phys. Rev.} {\bf D100} (2019), no.~3
  035041, [\href{http://arxiv.org/abs/1905.06344}{{\tt arXiv:1905.06344}}].

\bibitem{Han:2020oet}
C.~Han, M.~López-Ibáñez, B.~Peng, and J.~M. Yang, {\it {Dirac dark matter in
  $U(1)_{B-L}$ with Stueckelberg mechanism}},
  \href{http://arxiv.org/abs/2001.04078}{{\tt arXiv:2001.04078}}.

\bibitem{Luo:2020sho}
X.~Luo, W.~Rodejohann, and X.-J. Xu, {\it {Dirac neutrinos and $N_{{\rm
  eff}}$}},  {\em JCAP} {\bf 06} (2020) 058,
  [\href{http://arxiv.org/abs/2005.01629}{{\tt arXiv:2005.01629}}].

\bibitem{Borah:2020boy}
D.~Borah, A.~Dasgupta, C.~Majumdar, and D.~Nanda, {\it {Observing left-right
  symmetry in the cosmic microwave background}},  {\em Phys. Rev. D} {\bf 102}
  (2020), no.~3 035025, [\href{http://arxiv.org/abs/2005.02343}{{\tt
  arXiv:2005.02343}}].

\bibitem{Adshead:2020ekg}
P.~Adshead, Y.~Cui, A.~J. Long, and M.~Shamma, {\it {Unraveling the Dirac
  Neutrino with Cosmological and Terrestrial Detectors}},
  \href{http://arxiv.org/abs/2009.07852}{{\tt arXiv:2009.07852}}.

\bibitem{Luo:2020fdt}
X.~Luo, W.~Rodejohann, and X.-J. Xu, {\it {Dirac neutrinos and $N_{{\rm eff}}$
  II: the freeze-in case}},  \href{http://arxiv.org/abs/2011.13059}{{\tt
  arXiv:2011.13059}}.

\bibitem{Mahanta:2021plx}
D.~Mahanta and D.~Borah, {\it {Low scale Dirac leptogenesis and dark matter
  with observable $\Delta N_{\rm eff}$}},
  \href{http://arxiv.org/abs/2101.02092}{{\tt arXiv:2101.02092}}.

\bibitem{Du:2021idh}
Y.~Du and J.-H. Yu, {\it {Neutrino non-standard interactions meet precision
  measurements of $N_{\rm eff}$}},  \href{http://arxiv.org/abs/2101.10475}{{\tt
  arXiv:2101.10475}}.

\bibitem{Biswas:2021kio}
A.~Biswas, D.~Borah, and D.~Nanda, {\it {Light Dirac neutrino portal dark
  matter with observable $\Delta{N_{\rm eff}}$}},
  \href{http://arxiv.org/abs/2103.05648}{{\tt arXiv:2103.05648}}.

\bibitem{Li:2022yna}
S.-P. Li, X.-Q. Li, X.-S. Yan, and Y.-D. Yang, {\it {Cosmological imprints of
  Dirac neutrinos in a keV-vacuum 2HDM*}},  {\em Chin. Phys. C} {\bf 47}
  (2023), no.~4 043109, [\href{http://arxiv.org/abs/2202.10250}{{\tt
  arXiv:2202.10250}}].

\bibitem{Das:2023yhv}
N.~Das and D.~Borah, {\it {Light Dirac neutrino portal dark matter with gauged
  U(1)B-L symmetry}},  {\em Phys. Rev. D} {\bf 109} (2024), no.~7 075045,
  [\href{http://arxiv.org/abs/2312.06777}{{\tt arXiv:2312.06777}}].

\bibitem{Biswas:2022vkq}
A.~Biswas, D.~Borah, N.~Das, and D.~Nanda, {\it {Freeze-in dark matter via a
  light Dirac neutrino portal}},  {\em Phys. Rev. D} {\bf 107} (2023), no.~1
  015015, [\href{http://arxiv.org/abs/2205.01144}{{\tt arXiv:2205.01144}}].

\bibitem{Kuno:1999jp}
Y.~Kuno and Y.~Okada, {\it {Muon decay and physics beyond the standard model}},
   {\em Rev. Mod. Phys.} {\bf 73} (2001) 151--202,
  [\href{http://arxiv.org/abs/hep-ph/9909265}{{\tt hep-ph/9909265}}].

\bibitem{MEGII:2023ltw}
{\bf MEG II} Collaboration, K.~Afanaciev et~al., {\it {A search for $\mu^+ \rightarrow e^+ \gamma $ with the first dataset of the MEG~II
  experiment}},  {\em Eur. Phys. J. C} {\bf 84} (2024), no.~3 216,
  [\href{http://arxiv.org/abs/2310.12614}{{\tt arXiv:2310.12614}}].

\bibitem{ATLAS:2022yvh}
{\bf ATLAS} Collaboration, G.~Aad et~al., {\it {Search for invisible
  Higgs-boson decays in events with vector-boson fusion signatures using 139
  fb$^{-1}$ of proton-proton data recorded by the ATLAS experiment}},  {\em
  JHEP} {\bf 08} (2022) 104, [\href{http://arxiv.org/abs/2202.07953}{{\tt
  arXiv:2202.07953}}].

\bibitem{Gustafsson:2012aj}
M.~Gustafsson, S.~Rydbeck, L.~Lopez-Honorez, and E.~Lundstrom, {\it {Status of
  the Inert Doublet Model and the Role of multileptons at the LHC}},  {\em
  Phys. Rev.} {\bf D86} (2012) 075019,
  [\href{http://arxiv.org/abs/1206.6316}{{\tt arXiv:1206.6316}}].

\bibitem{Datta:2016nfz}
A.~Datta, N.~Ganguly, N.~Khan, and S.~Rakshit, {\it {Exploring collider
  signatures of the inert Higgs doublet model}},  {\em Phys. Rev.} {\bf D95}
  (2017), no.~1 015017, [\href{http://arxiv.org/abs/1610.00648}{{\tt
  arXiv:1610.00648}}].

\bibitem{Poulose:2016lvz}
P.~Poulose, S.~Sahoo, and K.~Sridhar, {\it {Exploring the Inert Doublet Model
  through the dijet plus missing transverse energy channel at the LHC}},  {\em
  Phys. Lett.} {\bf B765} (2017) 300--306,
  [\href{http://arxiv.org/abs/1604.03045}{{\tt arXiv:1604.03045}}].

\bibitem{Hashemi:2016wup}
M.~Hashemi and S.~Najjari, {\it {Observability of Inert Scalars at the LHC}},
  \href{http://arxiv.org/abs/1611.07827}{{\tt arXiv:1611.07827}}.

\bibitem{Belyaev:2016lok}
A.~Belyaev, G.~Cacciapaglia, I.~P. Ivanov, F.~Rojas, and M.~Thomas, {\it
  {Anatomy of the Inert Two Higgs Doublet Model in the light of the LHC and
  non-LHC Dark Matter Searches}},  \href{http://arxiv.org/abs/1612.00511}{{\tt
  arXiv:1612.00511}}.

\bibitem{Belyaev:2018ext}
A.~Belyaev, T.~R. Fernandez Perez~Tomei, P.~G. Mercadante, C.~S. Moon,
  S.~Moretti, S.~F. Novaes, L.~Panizzi, F.~Rojas, and M.~Thomas, {\it
  {Advancing LHC probes of dark matter from the inert two-Higgs-doublet model
  with the monojet signal}},  {\em Phys. Rev. D} {\bf 99} (2019), no.~1 015011,
  [\href{http://arxiv.org/abs/1809.00933}{{\tt arXiv:1809.00933}}].

\bibitem{CMS:2021kom}
{\bf CMS} Collaboration, A.~M. Sirunyan et~al., {\it {Measurements of Higgs
  boson production cross sections and couplings in the diphoton decay channel
  at $ \sqrt{\mathrm{s}} $ = 13 TeV}},  {\em JHEP} {\bf 07} (2021) 027,
  [\href{http://arxiv.org/abs/2103.06956}{{\tt arXiv:2103.06956}}].

\bibitem{Barman:2022yos}
B.~Barman, D.~Borah, A.~Dasgupta, and A.~Ghoshal, {\it {Probing high scale
  Dirac leptogenesis via gravitational waves from domain walls}},  {\em Phys.
  Rev. D} {\bf 106} (2022), no.~1 015007,
  [\href{http://arxiv.org/abs/2205.03422}{{\tt arXiv:2205.03422}}].

\bibitem{Barman:2023fad}
B.~Barman, D.~Borah, S.~Jyoti~Das, and I.~Saha, {\it {Scale of Dirac
  leptogenesis and left-right symmetry in the light of recent PTA results}},
  {\em JCAP} {\bf 10} (2023) 053, [\href{http://arxiv.org/abs/2307.00656}{{\tt
  arXiv:2307.00656}}].

\bibitem{Gelmini:2020bqg}
G.~B. Gelmini, S.~Pascoli, E.~Vitagliano, and Y.-L. Zhou, {\it {Gravitational
  wave signatures from discrete flavor symmetries}},  {\em JCAP} {\bf 02}
  (2021) 032, [\href{http://arxiv.org/abs/2009.01903}{{\tt arXiv:2009.01903}}].

\bibitem{King:2023ayw}
S.~F. King, R.~Roshan, X.~Wang, G.~White, and M.~Yamazaki, {\it {Quantum
  gravity effects on dark matter and gravitational waves}},  {\em Phys. Rev. D}
  {\bf 109} (2024), no.~2 024057, [\href{http://arxiv.org/abs/2308.03724}{{\tt
  arXiv:2308.03724}}].

\bibitem{Borah:2024kfn}
D.~Borah, N.~Das, and R.~Roshan, {\it {Testing quantum gravity effects with
  global lepton number symmetry and dark matter}},
  \href{http://arxiv.org/abs/2406.04404}{{\tt arXiv:2406.04404}}.

\bibitem{He:2006dk}
X.-G. He, Y.-Y. Keum, and R.~R. Volkas, {\it {A(4) flavor symmetry breaking
  scheme for understanding quark and neutrino mixing angles}},  {\em JHEP} {\bf
  04} (2006) 039, [\href{http://arxiv.org/abs/hep-ph/0601001}{{\tt
  hep-ph/0601001}}].

\bibitem{Ishimori:2010au}
H.~Ishimori, T.~Kobayashi, H.~Ohki, Y.~Shimizu, H.~Okada, and M.~Tanimoto, {\it
  {Non-Abelian Discrete Symmetries in Particle Physics}},  {\em Prog. Theor.
  Phys. Suppl.} {\bf 183} (2010) 1--163,
  [\href{http://arxiv.org/abs/1003.3552}{{\tt arXiv:1003.3552}}].

\end{thebibliography}
\providecommand{\href}[2]{#2}\begingroup\raggedright\endgroup

\end{document}